\documentclass[hidelinks,12pt]{article}
\usepackage{hyperref,natbib,amsmath,amssymb, amsthm}
\usepackage{import, color,graphicx}
\usepackage{soul}
\usepackage{comment}

\usepackage{diagbox}

\newcommand{\blu}[1]{\textcolor{black}{#1}} 

\usepackage{algpseudocode,algorithm}
\usepackage{hyperref,natbib,amsmath,amssymb, amsthm}
\usepackage{import, color,graphicx}
\usepackage{xcolor}

\usepackage{epsf}
\usepackage{epsfig}
\usepackage{amsfonts}
\usepackage{setspace}
\usepackage{float}
\usepackage{mathtools}
\usepackage{enumitem}

\DeclareMathOperator{\tr}{tr}

\newcommand{\bE}[0]{\mathbb{E}}

\newcommand{\x}{\mathbf{x}}
\newcommand{\xu}{\bar{\mathbf{x}}} 

\newcommand{\vecX}{\mathbf{X}}
\newcommand{\vecXu}{{\bar{\mathbf{X}}_n}} 
\newcommand{\vecY}{\mathbf{Y}}
\newcommand{\A}{\mathbf{A}}
\newcommand{\An}{\mathbf{A}_n}

\newcommand{\Cdelta}{\mathbf{C_{(\boldsymbol{\delta})}}}

\newcommand{\CN}{\mathbf{C}_N}

\newcommand{\Deltan}{\boldsymbol{\Delta}_n}

\newcommand{\K}{\mathbf{K}}
\newcommand{\KN}{\mathbf{K}_N}
\newcommand{\Kn}{\mathbf{K}_n}

\newcommand{\Lan}{\boldsymbol{\Lambda}_n}
\newcommand{\LaN}{\boldsymbol{\Lambda}_N}

\newcommand{\xnew}{\widetilde{\vecX}}
\newcommand{\vecxnew}{\tilde{\x}}
\newcommand{\xnewip}{\tilde{\x}_{i(p)}}

\newcommand{\cs}{\check{\sigma}}

\newcommand{\W}{\mathbf{W}}
\newcommand{\Wn}{\mathbf{W}_n}

\newcommand{\erf}{\mathrm{erf}}

\newcommand{\blind}{0}

\def\spacingset#1{\renewcommand{\baselinestretch}%
	{#1}\small\normalsize} \spacingset{1}

\begin{document}

	\if0\blind
	{
		\title{\vspace{-1cm}  Batch-sequential design and heteroskedastic surrogate modeling for
			delta smelt conservation}
		\author{
			Boya Zhang\thanks{Corresponding author: Department of Statistics, Virginia Tech, \href{mailto:boya66@vt.edu}{\tt boya66@vt.edu}}\
			\and 
			Robert B.~Gramacy\thanks{Department of Statistics, Virginia Tech}
			\and
			Leah Johnson\footnotemark[2]
			\and
			Kenneth A. Rose\thanks{University of Maryland Center for Environmental Science, Horn Point Laboratory, Cambridge, MD}
			\and 
			Eric Smith\footnotemark[2]
		}
		\date{}
		\maketitle
	}\fi
	
	\if1\blind
	{
		\bigskip
		\bigskip
		\bigskip
		\begin{center}
			{\LARGE\bf }
		\end{center}
		\medskip	
		\bigskip
	} \fi
	
\begin{abstract} 
Delta smelt is an endangered fish species in the San Francisco estuary that
have shown an overall population decline over the past 30 years. Researchers
have developed a stochastic, agent-based simulator to virtualize the system,
with the goal of understanding the relative contribution of natural and
anthropogenic factors that might play a role in their decline. However, the
input configuration space is high-dimensional, running the simulator is
time-consuming, and its noisy outputs change nonlinearly in both mean and
variance.  Getting enough runs to effectively learn input--output dynamics
requires both a nimble modeling strategy and parallel evaluation. Recent
advances in heteroskedastic Gaussian process (HetGP) surrogate modeling helps,
but little is known about how to appropriately plan experiments for highly
distributed simulation.  We propose a batch sequential design scheme,
generalizing one-at-a-time variance-based active learning for HetGP, as a
means of keeping multi-core cluster nodes fully engaged with runs. Our
acquisition strategy is carefully engineered to favor selection of replicates
which boost statistical and computational efficiency when training surrogates
to isolate signal from noise.  Design and modeling are illustrated on a range
of toy examples before embarking on a large-scale smelt simulation campaign
and downstream high-fidelity input sensitivity analysis.
	
\end{abstract}
	
	\if0\blind
	\fi
	
\section{Introduction}

Delta smelt is a short-lived fish species that spends its entire life within
the San Francisco Estuary (SFE) that connects the Sacramento and San
Joaquin Rivers through the Bay into the Pacific Ocean. The SFE has undergone
many changes over the past 150 years due to human development. It is now a
network of channels and sloughs surrounding islands protected by a man-made
levee system \citep{lund2010comparing}. 
Since 1960, environmental and ecological conditions (including delta smelt populations) have been extensively monitored. Derived population indices for delta smelt have shown a general decline since about
1980. In 1993 they were listed as threatened
under the US Endangered Species Act. The population exhibited a large drop in 2000 and has remained at low abundance \citep{stompe2020comparing}.  Factors that may contribute to
the decline of delta smelt include entrainment by water diversion
facilities, changes in food base and predation pressure, 
pollution, and changes in habitat related to salinity and turbidity
\citep{baxter2015updated, moyle2016delta}.

Identifying the relative importance of factors that impact the delta smelt
population is important for designing effective management actions to balance
human water use and maintenance/recovery of the species. Diverse statistical/observational
analyses have been entertained
\citep{Thomson:2010,MacNally:2010,miller2012investigation,MaunderandDeriso:2011,Hamilton:2018}.
%
\cite{kenny:2013,kenny:2013_2} developed a spatially-explicit, agent-based
model (ABM) of delta smelt to examine such factors. The model simulates daily
growth, mortality, reproduction, and movement of hundreds of thousands of
agents (smelt) from their birth to death. By explicitly representing food
(zooplankton), temperature, salinity, and water velocities experienced by
agents based on their location within a hydrodynamics grid, the ABM attempts
to scale local environmental effects up to population-level responses.
Simulations under myriad environmental and system variables enabled the
authors to identify conditions both detrimental and conducive to smelt growth
and survival, and to compare how changes in food and in entrainment from water
diversion facilities affect population growth rates \citep{Kenny:2018}.
Ultimately, the goal of such simulation is to augment and complement
statistical models, and to assist in determining which environmental factors
could affect the population of delta smelt.

\cite{kenny:2013,kenny:2013_2}'s ABM is coded in {\sf Fortran} and the version
we use takes about six hours to run. Even after pre-selecting a subset of key
model parameters, the input configuration space is large (upwards of
13-dimensions), and the outcome of the simulator varies across random seeds.
Response surface learning with this stochastic computer model -- separating
signal from noise in a high dimensional space -- requires a large, and costly,
distributed HPC simulation campaign and pairing with a flexible meta model. In
our initial study, described in Section \ref{sec:pilot}, we observe that the
response surface is nonlinear and heteroskedastic, i.e., sensitivity to
stochastic simulation dynamics is not uniform in the input space. These
features challenge effective meta-modeling, which are essential for downstream
tasks like input sensitivity analysis, mirroring ones which are increasingly
common the analysis of stochastic simulation experiments
\citep{baker2020stochastic}.

In similar but simpler situations
\citep[e.g.,][]{Johnson:2008,bisset2009epifast,farah2014bayesian,fadikar2018calibrating,rutter2019microsimulation}
-- being not as extreme as the delta smelt ABM in terms of simulator cost, input dimension, and
changing variance -- researchers have been getting mileage out of methods for
surrogate modeling and the design and analysis of computer experiments
\citep{Sacks:1989,santner2018design,gramacy2020surrogates}. Default,
model-free design strategies, such as space-filling Latin
hypercube samples \citep[LHS;][]{Mckay:1979}, are a good starting point but
are not reactive/easily refined to target parts of the input space which
require heavier sampling.  Model-based designs based on Gaussian process (GP)
surrogates fare better, in part because they can be developed sequentially
along with learning
\citep[e.g.,][]{jones:schonlau:welch:1998,seo00,gramacy2011particle}.

Until recently, surrogate modeling and computer experiment design methodology
has emphasized deterministic computer evaluations, for example those arising
in finite element analysis or solving systems of differential equations.
Sequential design with heteroskedastic GP (HetGP) surrogates
\citep{mickael:prac} for stochastic simulations has recently been proposed as
a means of dynamically allocating more runs in higher uncertainty
regions of the input space \citep{HetGP2}. Such schemes are typically
applied as one-at-a-time affairs -- fit model, optimize acquisition criteria,
run simulation, augment data, repeat -- which would take too long for the delta
smelt model. We anticipate needing thousands of runs, with several hours per run.
That process cannot be fully serial.

Batch-sequential design procedures have been applied with GP surrogates
\citep[e.g.][]{loeppky2010batch,ginsbourger2010kriging,chevalier2013fast,Duan:2017,Erickson:2018}.
These attempt to calculate a group of runs to go at once, say on a multi-core
supercomputing node, towards various design goals.  
Quasi-batch schemes, which asynchronously
re-order points for an unknown number of future simulations have also thrived
in HPC environments \citep{gra:lee:2009,taddy2009bayesian}.  However,
none of these schemes explicitly address input-dependent noise like we observe
in the delta smelt ABM simulations. Here we propose extending the
one-at-a-time method of \citet{HetGP2} to a batch-sequential setting.  Our
goal is to design for batches of size 24 to match the number of cores
available on nodes of a supercomputing cluster at Virginia Tech.  Following
\citeauthor{HetGP2}'s lead, we develop a novel scheme for encouraging
replicates in the batches. Replication is a tried and true technique for
separating signal from noise, reducing sufficient statistics for modeling and
thus enhancing computational and learning efficiency.

Our flow is as follows. Section \ref{sec:smelt} reviews simulation, surrogate
modeling and design elements for delta smelt simulations.
We also describe a pilot study on a reduced input space identifying
challenges/appropriate modeling elements and motivating a HetGP framework.
Section \ref{sec:batch_seq} explains our innovative batch-sequential
acquisition strategy through an integrated mean-squared prediction error
(IMSPE) criteria and closed-form derivatives for optimization, extending the
one-at-a-time process from \citet{HetGP2}. Section \ref{sec:merging} provides
a novel and thrifty post-processing scheme to identify replicates in the new
batch. Illustrative examples are provided throughout, and Section
\ref{sec:examples} details a benchmarking exercise against the infeasible
one-at-a-time gold standard.  Finally, in Section \ref{sec:deltasmelt} we
apply the design method to smelt simulations (in a larger space), collecting
thousands of runs utilizing tens of thousands of core hours across a several
weeks-long simulation campaign.  Those runs are used to conduct a
sensitivity analysis to exemplify potential downstream tasks.  We
conclude with other suggestions and methodological ideas in Section
\ref{sec:futurework}.

\section{Problem description and solution elements}
\label{sec:smelt}

Here we describe the smelt simulator and HPC implementation,  review 
surrogate modeling elements and report on a pilot study, motivating our
methodological developments.

\subsection{Agent-based model}
\label{sec:abm}

The delta smelt population model of \citet{kenny:2013}
is a stochastic, spatially-explicit, agent-based model (ABM). It tracks 
reproduction, growth, mortality, and movement of individual fish through 
life stages. 
Agents move around within a 1d network of channels and nodes formed
by rivers and leveed islands. Daily values of environmental variables of water
temperature, salinity, and the densities of six zooplankton prey types drive
the model.  These vary daily and spatially among channels over a grid. In our
simulations, these drivers are based on observed environmental variables from
1995 to 2005 to allow exploration of potential factors influencing a
population decline starting around 2000.

The model assumes that factors impact the agents in specific ways. Temperature
and zooplankton affect daily growth, whereas hydrodynamic transport and
salinity affect movement. Daily mortality is comprised of stage-specific rates
(mostly predation), starvation, and entrainment of individuals by water
diversion facilities. Daily egg production is used to start agents as yolk-sac
larvae. 
Upon progressing through multiple life stages and reaching maturity,
individuals spawn a year later, and the cycle repeats year-after-year.
Stochastic calculations within a run include: realization of zooplankton
concentrations in channels each day from regional means, assignment of
temperature of spawning to adults, aspects of hourly water transport of larvae
and the twice-per-day movement of juveniles and adults, timing of upstream and
downstream spawning migration, and selection of channels when individuals are
moved out of nodes (reservoirs).

\begin{table}[h!]
\centering
	\begin{tabular}{l l l c c c}
	\hline
	symbol & parameter & description &range   &default &pilot study\\
	\hline \hline
	$m_y$ & zmorty & yolk-sac larva MR 
	&[0.01,0.50]  &0.035 &0.035\\
	$m_l$ & zmortl & larval MR     &[0.01, 0.08]   &0.050 &0.050\\
	$m_p$ & zmortp & post-larval MR &[0.005, 0.05]	 &0.030 &0.030\\
	$m_j$ & zmortj & juvenile MR   &[0.001, 0.025] &0.015 &[0.005,0.030]\\
	$m_a$ &zmorta & adult MR     &[0.001, 0.01]  &0.006 &0.006\\
	\hline
	$m_r$ & middlemort & river entrain MR  &[0.005, 0.05]	&0.020 &[0, 0.05]\\
	\hline
	$P_{l,2}$ & preyk(3,2)& larvae EPT 2         &[0.10, 20.0]	   &0.200 &0.200\\
	$P_{p,2}$ & preyk(4,2)& postlarvae EPT 2    &[0.10, 20.0]      &0.800 &[0.10, 1.84]\\ 
	$P_{p,6}$ & preyk(4,6)& postlarvae EPT 6    &[0.10, 20.0]	   &1.500 & $P_{p,2}$\\
	$P_{j,3}$ & preyk(5,3)& juveniles EPT 3      &[0.10, 20.0]	  &0.600 &[0.1, 1.5]\\ 
	$P_{j,6}$ & preyk(5,6)& juveniles EPT 6      &[0.10, 20.0]	   &0.600 &  $P_{j,3}$\\
	$P_{a,3}$ & preyk(6,3)& adults EPT 3         &[0.01, 20.0]	   &0.070 &0.070\\
	$P_{a,4}$ & preyk(6,4)& adults EPT 4         &[0.01, 5.0] 	   &0.070 &0.070\\
	\hline	
	\end{tabular}
	\caption{Smelt simulator inputs considered in this analysis.
	The pilot study column indicates settings for Section \ref{sec:pilot}. MR is
	mortality rate; EPT means eating prey type.}
	\label{tab:deltasmelt}
\end{table}

The Rose et al.~ABM has more than fifty model parameters. 
We selected 13 of these to focus on in this analysis because they are known to
be important to model dynamics and have direct relevance to the ecology and
management of delta smelt.  They are listed in Table
\ref{tab:deltasmelt} with their symbols and input parameter name,
descriptions, range extremes, default/calibrated values, and spans considered
in our pilot study (Section \ref{sec:pilot}).

The first set of parameters in Table \ref{tab:deltasmelt} involve
natural mortality rates assigned to each life stage.
While there is uncertainty in these values, feasible ranges can be deduced from prior analysis and review of values reported in the literature. 
The second group is a single parameter ($m_r$) that modifies the 
mortality rate of juvenile and adults 
based on whether
river flows in the Delta subregion are transporting individual fish toward or
away from water diversion facilities. Flows towards facilities result in
the addition of $m_r$ due to entrainment.

The third group of parameters are feeding-related and are specific to prey
group and life stage; for example, $P_{j,6}$ is juveniles feeding on prey type
6 ({\em Pseudodiaptomus forbesi}).  The parameters are half-saturation
coefficients in a functional response feeding relationship and so larger
values reduce feeding rates. Prey types are selected for each life stage that
were dominant in simulated diets for the time period analyzed here
\citep{kenny:2013}.  Dependencies are created among feeding parameters to
mimic the effects of more or less food available to life stages that consume
multiple prey types. For example, when $P_{p,2}$ is varied its value is pegged
to $P_{p,6}$, i.e., more or less food for post-larval stage. These
dependencies are noted in Table \ref{tab:deltasmelt} for the pilot study and
Table \ref{tab:smeltfullstudy} for the full analysis.

 A distinct feature of the \cite{kenny:2013} ABM is how model behavior is
 summarized.  Output is extensive because the model generates size (length and
 weight), location, growth rate, mortality from different sources, diet, and
 other individual-level attributes every day for approximately 450,000 model
 individuals for the ten-year simulations analyzed here. \citet{kenny:2013} summarize these dynamics using
 the information on individuals to estimate a matrix projection model for each
 year, ultimately generating a population growth rate ($\lambda_i$) each year.
 Here, we use the geometric mean of growth rates from 1995 to 2004 as
 \blu{a convenient scalar} output summarizing results of the 10-year
 simulation: $\lambda = (\prod_{i=1995}^{2004} \lambda_i)^{1/10}\!$. The value
 of $\lambda$ is an indicator of the health of the population of delta smelt
 over the time period of the simulation and is directly interpretable. Values
 greater than one indicate population growth over the 10 years; values less
 than one indicate decline.

Previous simulation campaigns did not systematically vary all the parameters
simultaneously. For example, 
the importance of single
factors were estimated by evaluating population changes after structurally
eliminating that factor in the simulation(s) \citep{Kenny:2018}. This is
very different from a Saltelli-style/functional analysis of variance
\citep[e.g.,][Chapter
8.2]{saltelli2000sensitivity,oakley2004probabilistic,marrel2009calculations,gramacy2020surrogates}
favored by the computer surrogate modeling literature.  That and other
downstream applications require a meta-modeling design strategy in the face of
extreme computational demands and stochasticity over random seeds.

\subsection{Surrogate modeling}
\label{sec:hetGP_background}

We regard the delta smelt simulator as an unknown function $f: \mathbb{R}^d
\rightarrow \mathbb{R}$.  A meta-model $\hat{f}$ fit to evaluations
$(\mathbf{x}_i, y_i \sim f(\mathbf{x}_i))$, for $i=1,\dots,N$ is known as a
surrogate model or emulator \citep{gramacy2020surrogates}. The idea is that
fast $\hat{f}(\mathbf{x})$ could be used in lieu of slow/expensive
$f(\mathbf{x})$ for downstream applications like input sensitivity analysis.
Although there are many sensible choices, the canonical surrogate is based on
Gaussian processes (GPs).  If $f$ is deterministic ($y_i = f(\mathbf{x}_i)$),
this amounts to specifying a multivariate normal (MVN) for
$\vecY_N=(y_1, \dots, y_N)^\top$:  $\vecY_N
\sim \mathcal{N}_N(\mathbf{0}, \blu{ \tau^2} \CN)$.  
\blu{Defining $\mathbf{C}_N$ based on distance, 
	\begin{equation*}
	\CN^{ij} = c_{\boldsymbol{\theta}}(\x_i, \x_j) = 
	\exp\left\{-\sum_{k=1}^{d}\frac{(\x_{ik} - \x_{jk})^2}{\theta_k}\right\},
	\end{equation*}
provides smooth decay in function space when moving apart in $\x$-space and yields a predictive surface interpolates the data.}\footnote{Mat\`ern is another choice \citep{stein2012interpolation}; our contribution
is kernel agnostic as long it is differentiable.}
\blu{Fixing $\boldsymbol{\theta}$ and $\tau^2$, dropping $\boldsymbol{\theta}$ in 
	$c_{\boldsymbol{\theta}}(\cdot, \cdot)$,}
 extending the MVN to the cover $(N+N')$-sized $(\vecY_N,
\mathcal{Y}(\mathcal{X}))$ at training inputs $\vecX_N$ and $N'$ testing sites
$\mathcal{X}$, and MVN conditioning leads to a Gaussian predictive distribution
$\mathcal{Y}(\mathcal{X})
\mid \mathbf{Y}_N$ with
\begin{align}
\mbox{mean} && \mu(\mathcal{X} \mid \vecY_N) &= c(\mathcal{X}, \vecX_N)  \CN^{-1}\vecY_N,
\label{eq:predgp} \\
\mbox{and variance} &&
\Sigma(\mathcal{X} \mid \vecY_N) &= \tau^{2}[c(\mathcal{X},\mathcal{X})
- c(\mathcal{X}, \vecX_N)\CN^{-1} c(\mathcal{X}, \vecX_N)^\top]. \nonumber
\end{align}
Observe that uncertainty
$\Sigma(\mathcal{X} \mid \mathbf{Y}_N)$ is a quadratic function of 
\blu{distance between testing data $\mathcal{X}$ and training data $\vecX_N$ locations.}
For this reason, space-filling
designs such as maximin design \citep{Johnson:1990}, LHS and hybrids thereof like maximin--LHS
\citep{Morris:1995} are common in order to sufficiently cover the input space.

For stochastic $f$ with constant noise we can add a nugget term $g$ to the
diagonal $\KN = \CN + \LaN$ for $\LaN = g\mathbb{I}_N$ and take $\mathbf{Y}_N
\sim \mathcal{N}_N(0, \tau^2 \KN)$.  To model a response surface with
non-constant noise, \citet{mickael:prac} proposed freeing the diagonal
elements of $\LaN$ under a smoothness penalty.  They call this a
heteroskedastic GP (HetGP).  Specifically, let $\delta_1, \delta_2, \dots,
\delta_n$ denote latent nuggets, corresponding to $n \ll N$ unique design
locations.  Replication in a design, here with degree $N-n$, is essential for
separating signal from noise and also leads to computational efficiencies,
working with cubic in $n$ rather than $N$ flops through a Woodbury trick not
reviewed here.\footnote{It is also important not to introduce latent
$\delta_i$ in multitude at identical input locations $\mathbf{x}_i$ which
introduces numerical instabilities to the inferential scheme.} Place
\blu{$\delta_1, \delta_2, \dots, \delta_n$} diagonally in $\mathbf{\Delta}_n$
and assign to these 
a structure similar to $\mathbf{Y}$ but now encoding a prior on
variances: $
\Deltan \sim \mathcal{N}_n(\mathbf{0}, \tau^2_{(\boldsymbol{\delta})} (\Cdelta + g_{(\boldsymbol{\delta})} \An^{-1}))$.
$\Cdelta$ is the covariance matrix of $n$ unique design locations defined
under similar kernel/inverse distance structure; $\An$ is a diagonal matrix,
$\A_{ii} = a_i$, which denotes the number of replicates at unique location
$\bar{\mathbf{x}}_i$ so that $\sum_{i=1}^n a_i = N$; $g_{(\boldsymbol{\delta})}$ is
a ``nugget of nuggets'' controlling the smoothness of $\lambda_i$'s relative
to $\delta_i$'s.\footnote{This $\lambda_i$ notation, from \citet{hetGP1}, should not be confused with the delta smelt simulation output from \citet{kenny:2013}, whose logarithm we take as the main response ($y_i$) in our analysis.}  
 Smoothed $\lambda_i$-values can be calculated by plugging
$\Deltan$ into GP mean predictive equations \eqref{eq:predgp}:
\begin{equation}
	\Lan =  \Cdelta \K_{(\boldsymbol{\delta})}^{-1} \Deltan, \quad \mbox{ where  } \quad \K_{(\boldsymbol{\delta})} =  \Cdelta + g_{(\boldsymbol{\delta})} \An^{-1}.
	\label{eq:rhat}
\end{equation}
Parameters including $\boldsymbol{\theta}, \tau^2$ for both GPs, i.e., for
mean and variance, \blu{and latent nuggets $\Deltan$} may be estimated by
maximizing the joint log likelihood with derivatives  in time cubic in $n$.
Software is available for {\sf R} as {\tt hetGP} on CRAN \citep{mickael:prac}.

\subsection{Pilot study}
\label{sec:pilot}

To assist with {\sf R}-based surrogate modeling we built a custom {\sf R}
interface to the underlying {\sf Fortran} program automating the passing
of input configuration files/parsing of outputs through ordinary function I/O.
The {\tt Rmpi} package \citep{Rmpi} facilitates cluster-level parallel
evaluation for distributed simulation through a message passing interface
(MPI) on our Advanced Research Computing (ARC) HPC facility at Virginia Tech.

To test that interface and explore modeling and design options we ran a
limited delta smelt simulation campaign over six parameters (inputs)
under a maximin--LHS of size $n=96$ \citep[via {\tt lhs};][]{lhs} with five
replicates for each combination.  Juvenile and river entrainment mortalities
$m_j$ and $m_r$ were varied over their ranges with the rest of the
mortality rate parameters were fixed at their default values from Table
\ref{tab:deltasmelt}. Post-larvae $(P_{p,2})$ and juvenile $(P_{j,3})$ prey
parameters for zooplankton type 2 are allowed to vary over their ranges with
value of $P_{p,2}$ also being assigned to $P_{p,6}$ and value of $P_{j,3}$
also being assigned to $P_{j,6}$ ($P_{p,2}=P_{p,6}$ and $P_{j,3}=P_{j,6}$).
Other prey types were fixed to their default settings making the effective
input dimension four. Twenty 24-core VT/ARC cluster nodes were fully occupied
in parallel in order to get all $N=480$ runs in about six hours.

\begin{figure}[ht!]
	\centering
	\small
	\tabcolsep=0.11cm
	\begin{tabular}{ccc}
		\;\;\; mean surface & \;\; variance surface & \;\; 1d predictive bands \\
			
		\includegraphics[trim={10 25 0 25},width=52mm]{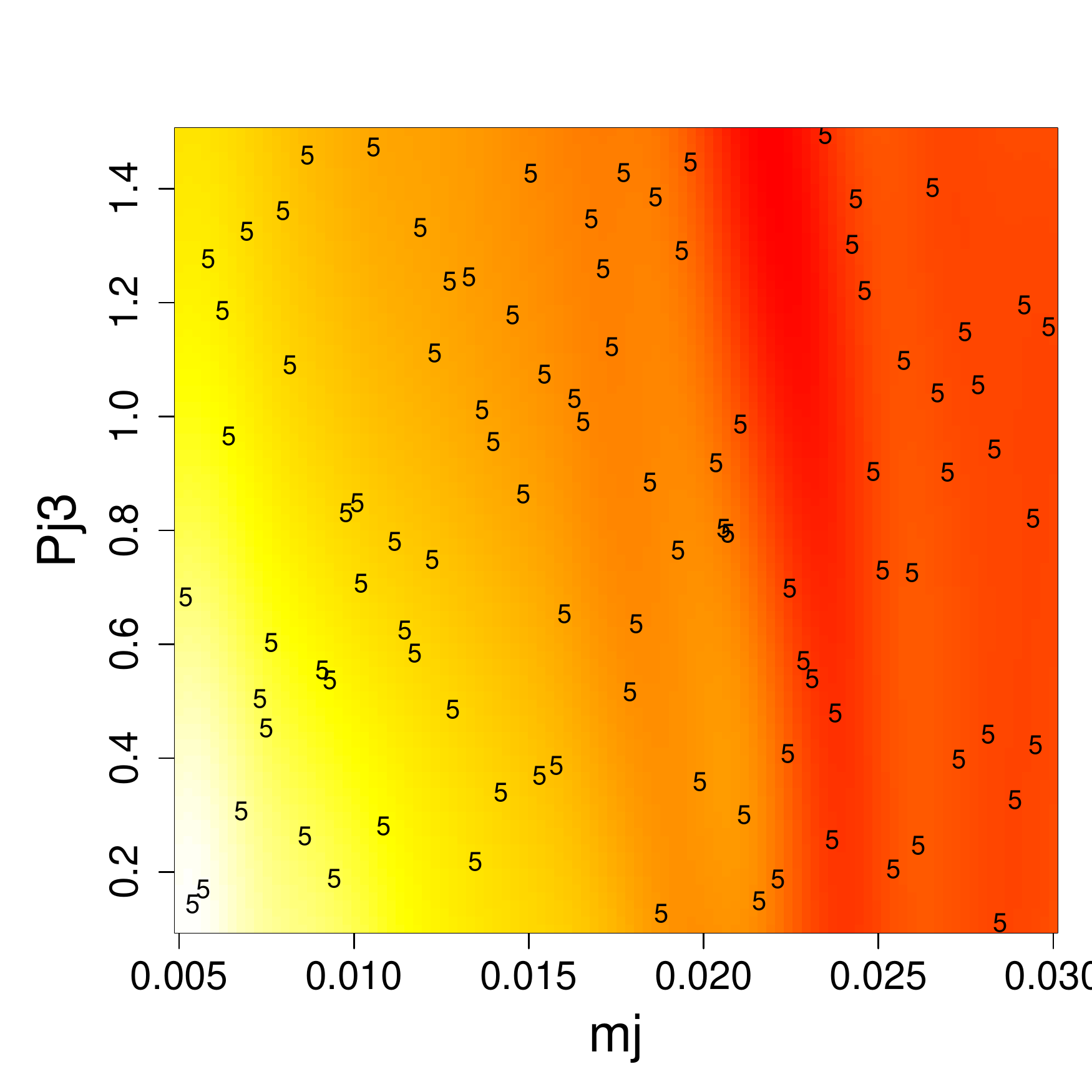} & 
		\includegraphics[trim={10 25 0 25},width=52mm]{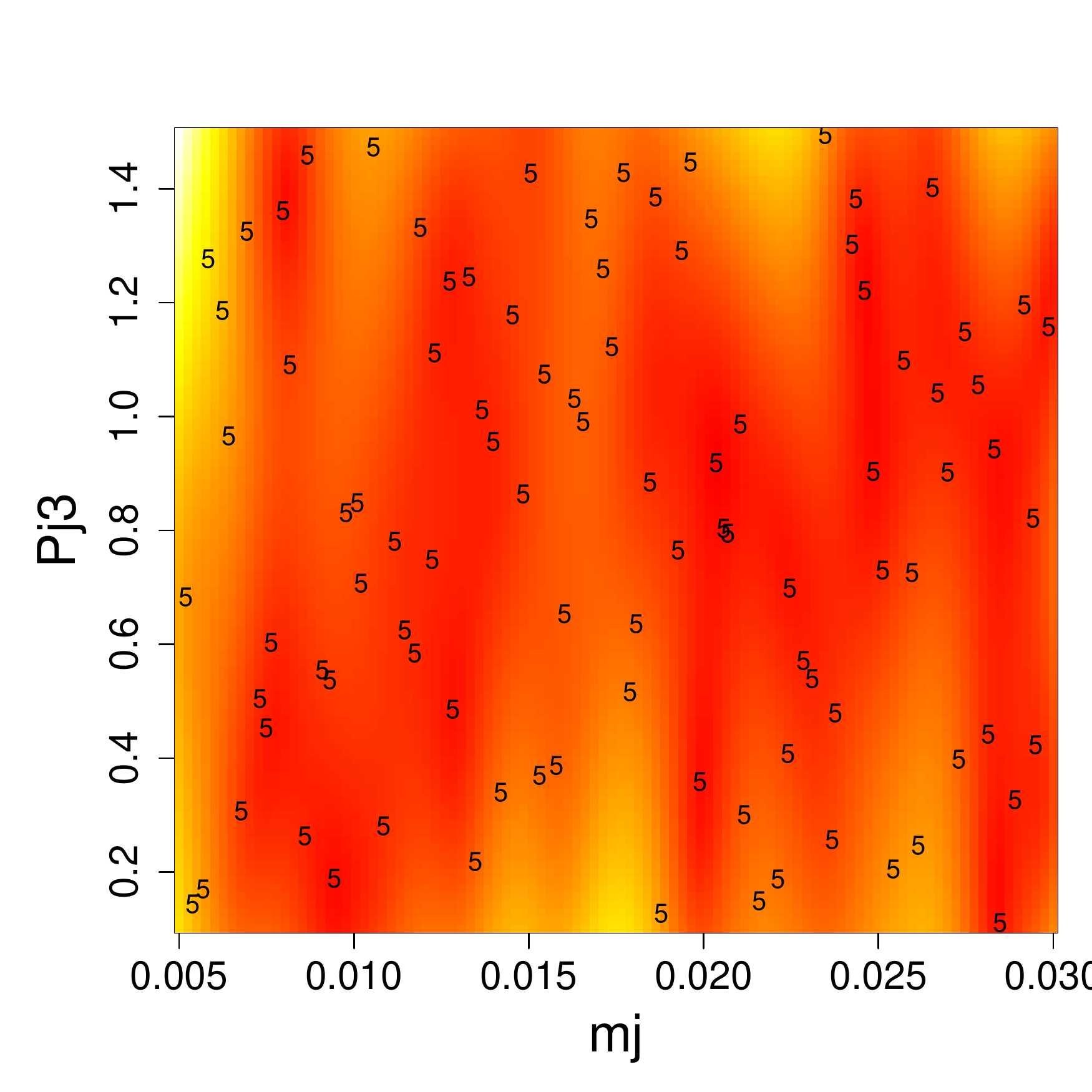} &  
		\includegraphics[trim={10 25 0 25},width=52mm]{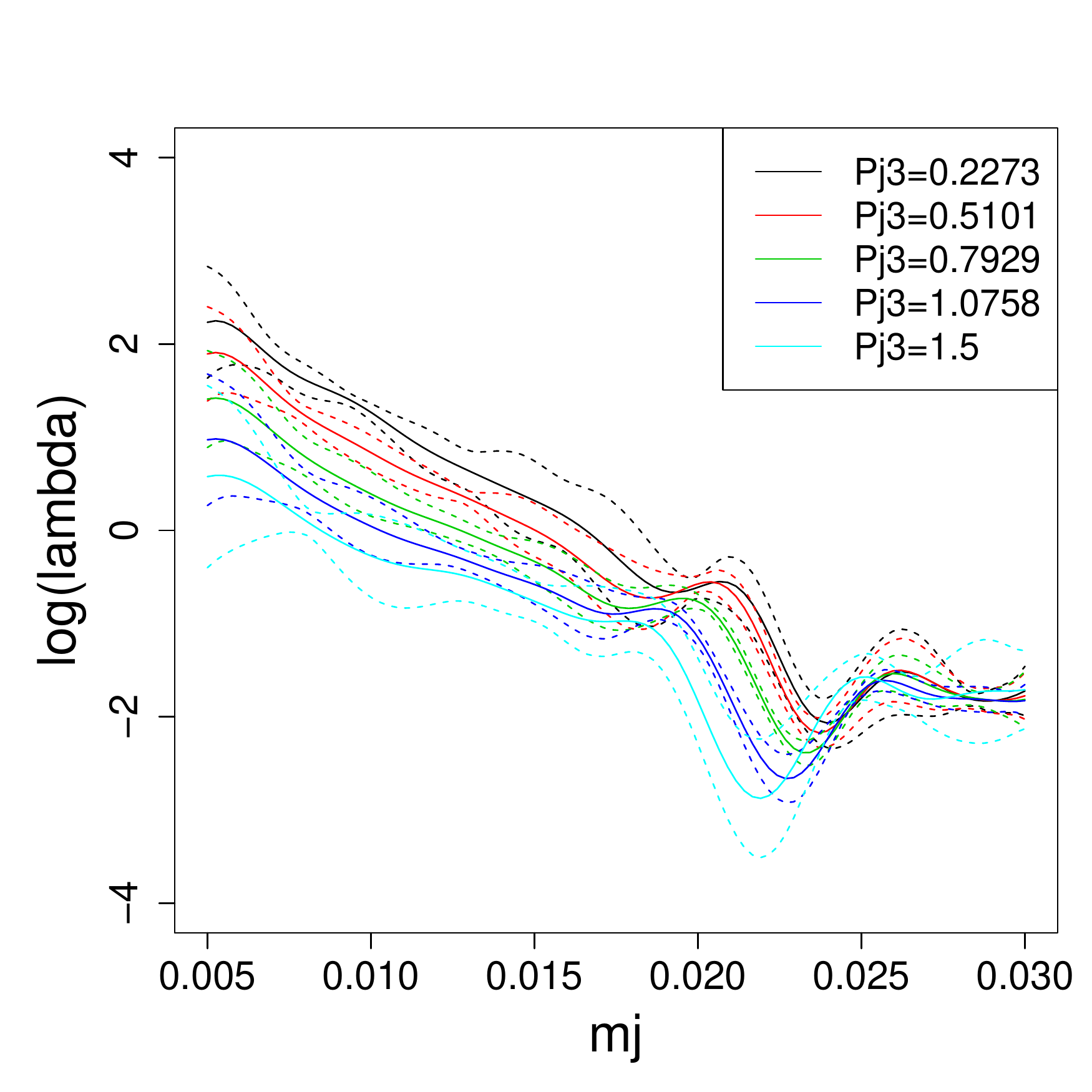} \\
		
		\includegraphics[trim={10 30 0 25},width=52mm]{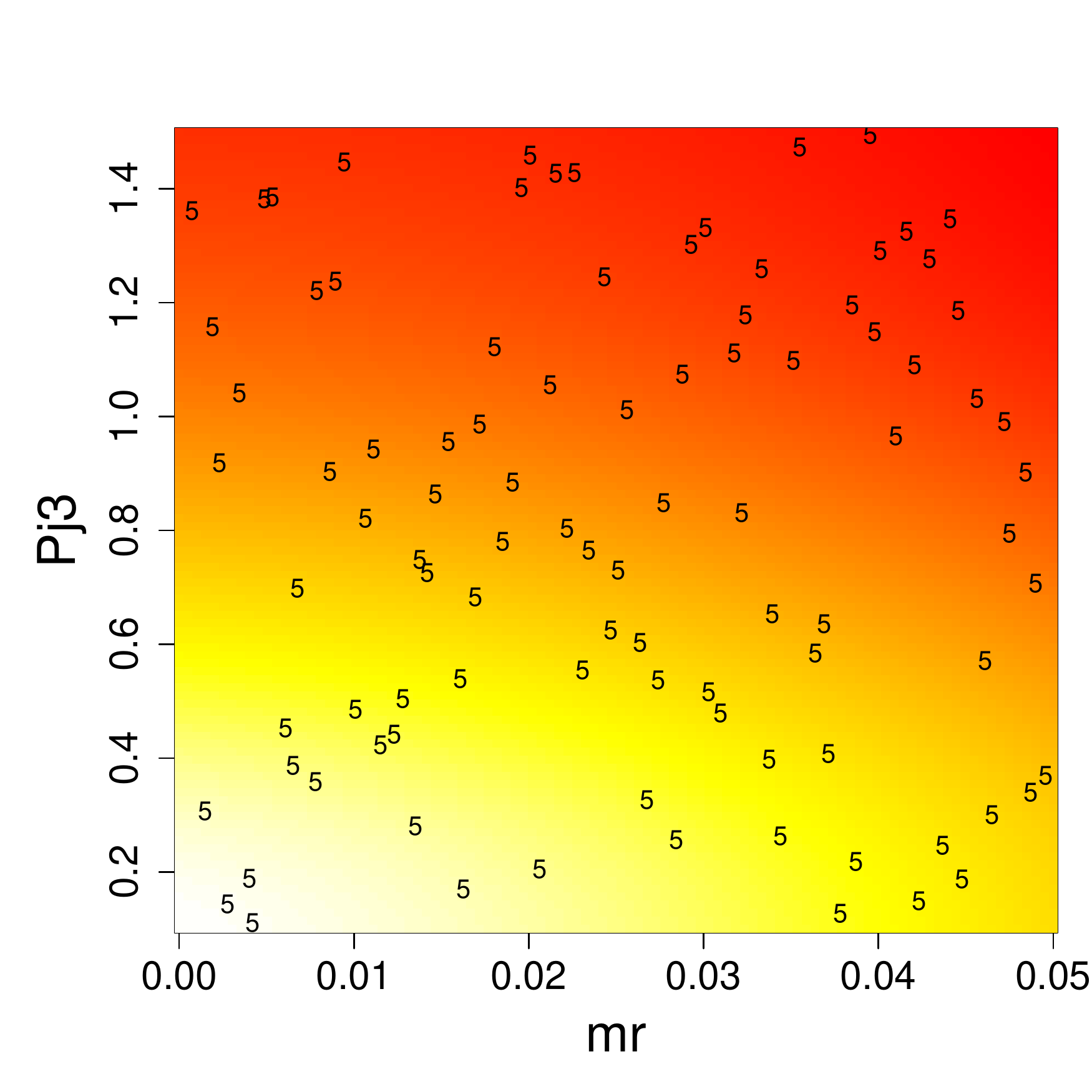} & 
		\includegraphics[trim={10 30 0 25},width=52mm]{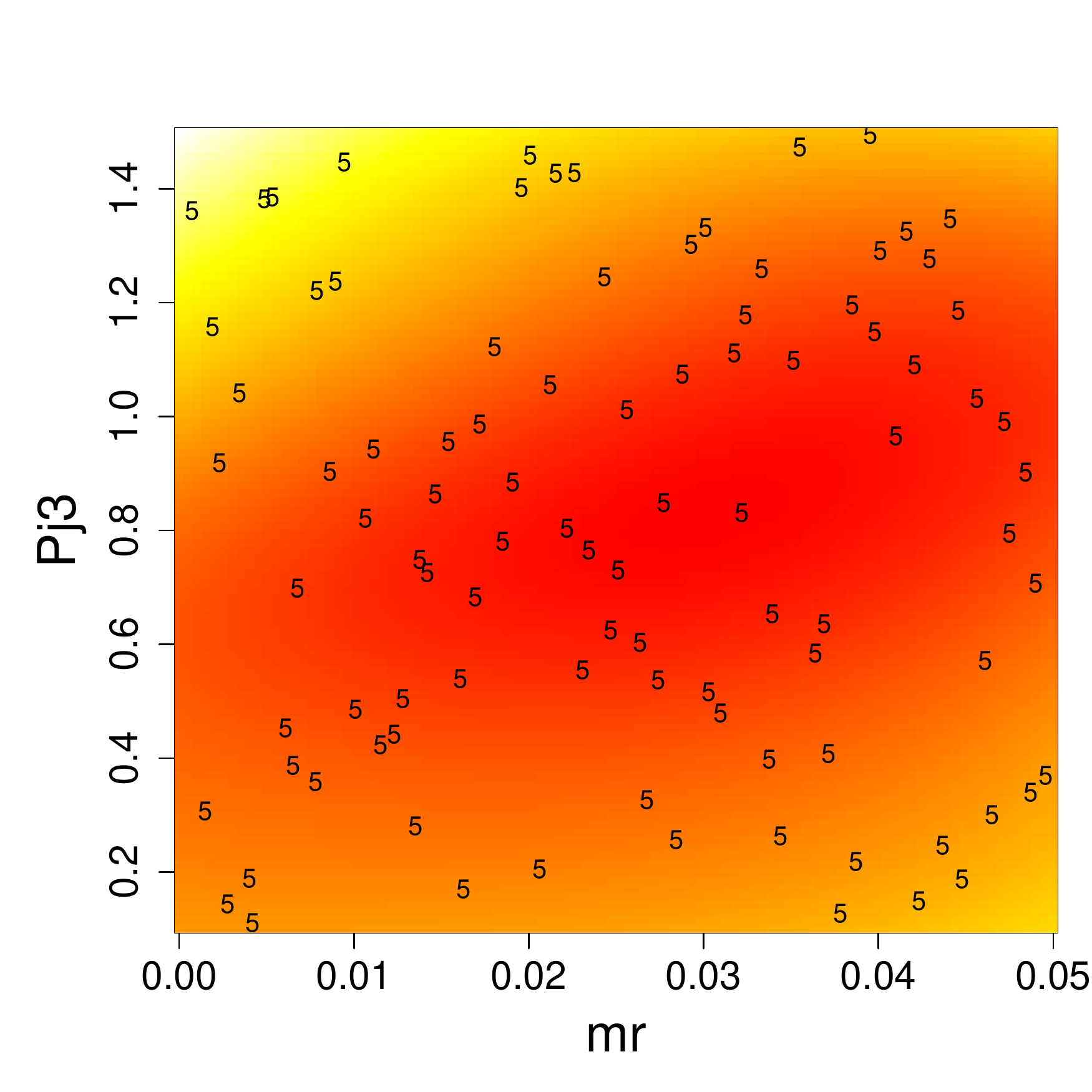} &  
		\includegraphics[trim={10 30 0 25},width=52mm]{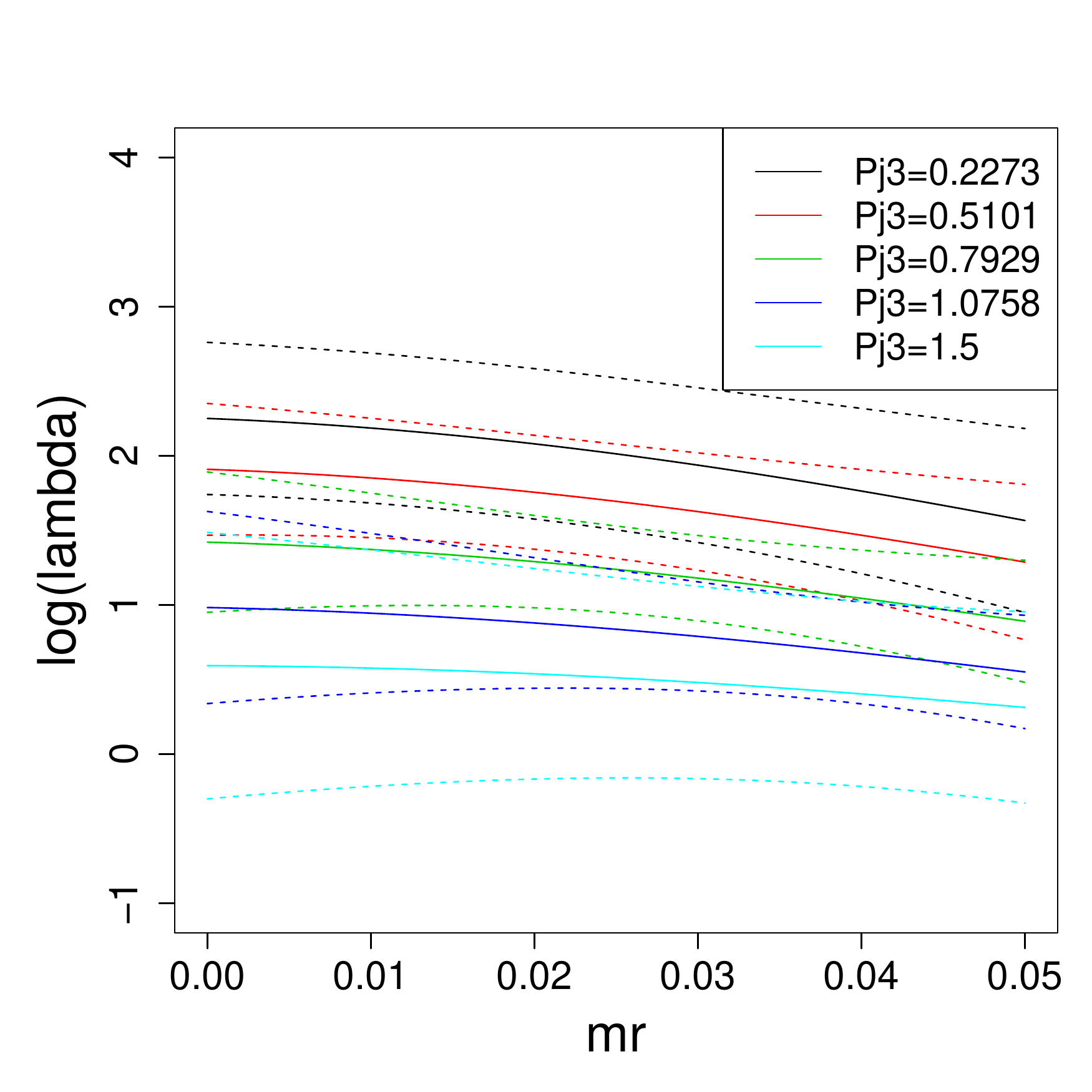}
	\end{tabular}
	\caption{2d heatmap and 1d lineplot slices of predictive mean and variance for selected inputs. The numbers overlaid indicate design locations and numbers of replicates.}
	\label{fig:screen}
\end{figure}

We fit the simulation data using {\tt hetGP} with inputs $\mathbf{X}_N$ coded
to the unit cube $[0,1]^4$ and with $\mathbf{Y}_N$ derived from $\log
\lambda_i$, the (log) 10-year geometric average of the population growth rate, for $i=1,\dots,480$ using $y_i
\equiv -6\log 10$ in the few cases where $\lambda_i = 0$ was returned.  As a
window into visualizing the fitted response surface we plotted a selection of
1d and 2d predictive mean error/variance slices in Figure \ref{fig:screen},
using defaults from Table \ref{tab:deltasmelt} for the fixed variables. The
first and second row correspond to subspaces $(P_{j,3}
\times m_j)$ and $(P_{j,3} \times m_r)$, respectively. Observe in the middle
column how noise intensity changes over the 2d input subspace, indicating
heteroskedasticity. Both mean and variance surfaces are nonlinear.  A similar,
higher resolution view is offered by the 1d slices in the final column. The
solid curves in the top-right panel are horizontal slices of the top left
panel with $P_{j,3}$ fixed at five different values, and analogously on the
bottom-right. Predictive intervals, \blu{with 95\% nominal coverage,} are
shown as dashed lines. In both views, the width of dashed predictive band
changes, sometimes drastically, as $m_j$ and $m_r$ are increased. Clearly
$m_j$ in the top-right panel shows more dramatic and nonlinear mean and
variance effects.
	
\section{Batch sequential design}
\label{sec:batch_seq}

The plan is to scale-up the pilot study of Section \ref{sec:pilot} and vary
more quantities in the 13d input space.  Ideally, sampling effort would
concentrate on parts of the input space that are harder to model, or where
more value can be extracted from noisy simulations. \citet{HetGP2} proposed
IMSPE-based one-at-a-time sequential design with that goal in mind. Here we
extend that to batches that can fill entire compute nodes at once.

\subsection{A criterion for minimizing variance}
\label{sec:imspe}

Integrated mean-squared prediction error (IMSPE) measures how well a surrogate
model captures input--output relationships. It is widely used as data
acquisition criterion; see, e.g., \citet[][Chapters 6 and
10]{gramacy2020surrogates}.  Let $\cs^2_N(\x)$ denote the nugget free
predictive variance for any single $\x \in D$. IMSPE for a design $\vecX_N$
may be defined as:
\[I_N \equiv \text{IMSPE}(\vecX_N) = \int \limits_{\x \in D}
\cs^2_N(\x) \, d\x=
\int \limits_{\x \in D}\hat{\tau}^{2}[c(\x, \x)
- c(\x, \vecX_N) \KN^{-1} c(\x, \vecX_N)^\top] \,d\x.
\]
The integral above has an analytic expression for GP surrogates, in part
because of the closed form for $\cs^2_N(\x)$. Examples involving
specialized GP setups in recent literature include
\citet{Ankenman2010,Leatherman2017,chen2019adaptive}. Similar
expressions do not, to our knowledge, exist for other popular surrogates like
deep neural networks, say.

\citet{HetGP2} gives perhaps the most generic and prescriptive expression for
GPs, emphasizing replicates at $n\ll N$ unique inputs $\bar{\mathbf{x}}_i$ for
computational efficiency.  Let $\K_n$ denote the unique $n\times n$ covariance
structure comprised of $\Kn^{ij} = c(\xu_i, \xu_j) + \delta_{ij}
\frac{r(\xu_i)}{a_i}$.  Let $\Wn$ be an $n
\times n$ matrix with entries comprising integrals of kernel products 
$w(\xu_i, \xu_j) =
	\int_{\x \in D} c(\xu_i, \x) c(\xu_j, \x) \; d\x$ for $1 \leq i,j \leq n$, and
let $E = \int_{\x \in D} \blu{\hat{\tau}^{2}}  c(\x,\x) \; d\x$, which is constant with respect to
the design $\vecX_n$. Closed forms are provided in Appendix B
of \citeauthor{HetGP2} for common kernels. Then $\mathcal{O}(n^3)$ calculations yield
\begin{equation}
	I_N = \bE[\blu{\hat{\tau}^{2}} c(X, X)] - \bE[ \blu{\hat{\tau}^{2}} c(X, \vecX_N)\KN^{-1} c(X, \vecX_N)^\top] = E - \blu{\hat{\tau}^{2}} \tr(\Kn^{-1}\Wn).
	\label{equ:IN}
\end{equation}

Although expressed for an entire design $\vecX_n$, in practice IMSPE is most
useful in sequential application where the goal is to choose new runs.
\citeauthor{HetGP2}~provided a tidy expression for solving for $\x_{n+1}$ by
optimizing $I_{n+1}(\tilde{\mathbf{x}})$ over $n+1^\mathrm{st}$ candidates
$\tilde{\mathbf{x}}$.  We extend this to an entire batch of size $M \geq 1$,
augmenting $\vecX_N$ or (more compactly) the unique elements $\bar{\vecX}_n$.
Let $\xnew =
\{\vecxnew_1, \vecxnew_2, \dots ,
\vecxnew_M\}^\top$ denote the coordinates of a new batch. Let $I_{N+M}(\xnew)$
denote the new IMSPE, which is realized most directly by shoving a
row-combined $[\vecX_N; \xnew]$ into Eq.~\eqref{equ:IN}.  That
over-simplifies, and flops in $\mathcal{O}((N+M)^3)$ could be
prohibitive.

Partition inverse equations \citep{barnett:1979} can be leveraged for even thriftier evaluation.  Extend the kernel $\K$ and its integral $\W$
to define new $(n+M)\times(n+M)$ matrices
\[
\K_{n+M}=\begin{bmatrix}
\K_{n} & c(\vecXu, \xnew) \\
c(\vecXu, \xnew) ^\top & c(\xnew,\xnew) + r(\xnew)
\end{bmatrix}, \quad
\W_{n+M}= \begin{bmatrix}
\Wn & w(\vecXu, \xnew)\\
w(\vecXu, \xnew)^\top & w(\xnew, \xnew)
\end{bmatrix},
\]
where $\Wn = w(\vecXu,
\vecXu)$ and $r(\xnew) = \mathrm{Diag}(r(\vecxnew_1), \dots, r(\vecxnew_M))$ 
comes from smoothed latent variances following
Eq.~\eqref{eq:rhat} via \blu{$c_{(\boldsymbol{\delta})}(\xnew, \vecXu)$} so that 
\begin{equation}
\blu{r}(\xnew) = \blu{c}_{(\boldsymbol{\delta})}(\xnew, \vecXu)(\Cdelta + g_{(\delta)} \An^{-1})^{-1} \Deltan. \label{equ:rhatpred}
\end{equation}
%
\blu{Given $\K_n^{-1}$}, we may fill the inverse $\K_{n+M}^{-1}$ in flops in $\mathcal{O}(M^3
+ nM^2 + n^2M)$ as
\begin{equation}
\K_{n+M}^{-1}=\begin{bmatrix}
\K_{n}^{-1}+ g(\xnew) \Sigma(\xnew) g(\xnew)^ \top & g(\xnew) \\
g(\xnew)^\top & \Sigma(\xnew)^{-1}
\end{bmatrix},
\label{eq:Knp1i_up}
\end{equation}
where $g(\xnew)= -\K_n^{-1} c(\vecX_n, \xnew)\Sigma(\xnew)^{-1}$,
$\Sigma(\xnew) = r(\xnew) + c(\xnew, \xnew) - c(\vecXu, \xnew)^\top
\K_{n}^{-1}c(\vecXu, \xnew)$.
Multiplying through components of Eq.~\eqref{eq:Knp1i_up} and properties of
traces in Eq.~\eqref{equ:IN} leads to
\begin{align}
I_{N+M} &= E - \blu{\hat{\tau}^{2}} \tr(\K_n^{-1} \Wn + g(\xnew)\Sigma(\xnew) g(\xnew)^{\top} 
+ g(\xnew) w(\vecX_n, \xnew)^{\top}) \nonumber \\
 &\quad - \blu{\hat{\tau}^{2}} \tr(g(\xnew)^{\top} w(\vecX_n, \xnew) + \Sigma(\xnew)^{-1} w(\xnew, \xnew)) \label{equ:Inm} \\
&= I_N - \blu{\hat{\tau}^{2}} \left[\tr(g(\xnew)\Sigma(\xnew) g(\xnew)^{\top}) + 2\tr(g(\xnew) w(\vecX_n, \xnew)^{\top}) + \tr(\Sigma(\xnew)^{-1} w(\xnew, \xnew))\right].
\nonumber
\end{align}
Finding the best $\xnew$ requires only the latter term above.  That is, we seek
\begin{align*}
\xnew^* 
&= \mathrm{argmax}_{\xnew \in D}
\tr(g(\xnew)\Sigma(\xnew) g(\xnew)^{\top}) + 2\tr(g(\xnew) w(\vecX_n, \xnew)^{\top}) + \tr(\Sigma(\xnew)^{-1} w(\xnew, \xnew)).
\end{align*}
In other words, we seek $\xnew^*$ giving the largest reduction in IMSPE.
Evaluation involves flops in the orders quoted above, however in
repeated calls for numerical optimization many of the $\mathcal{O}(n)$
quantities can be pre-evaluated leaving $\mathcal{O}(M^3
+ nM^2 + n^2M)$ for each $\xnew$. 

\subsection{Batch IMSPE gradient}
\label{sec:gradient}

To facilitate library based numerical optimization of $I_{N+M}(\xnew)$ with
respect to $\xnew$, in particular via Eq.~\eqref{equ:Inm}, we furnish closed-form
expressions for its gradient.  Below, these are framed via partial derivatives
for $\xnewip$, the $p^\mathrm{th}$ coordinate of the $i^\mathrm{th}$
subsequent design point in the new batch. Beginning with the chain rule,
the  gradient of $I_{N+M}$ over
$\xnewip$ follows
\begin{equation}
\frac{\partial I_{N+M}}{\partial \xnewip} = - \blu{\hat{\tau}^{2}}\mathrm{tr}\left(\frac{\partial \K_{n+M}^{-1}}{\partial \xnewip}\W_{n+M} + \K_{n+M}^{-1} \frac{\partial \W_{n+M}}{\partial \xnewip}\right).
\label{equ:chainrule}
\end{equation}
Recursing through its component parts, we have
\begin{align*}
\dfrac{ \partial \K^{-1}_{n+M}}{\partial \xnewip} & =\dfrac{ \partial} {\partial \xnewip} \begin{bmatrix}
\K_{n}^{-1}+ g(\xnew) \Sigma(\xnew) g(\xnew)^ \top & g(\xnew) \\
	g(\xnew) ^\top & \Sigma(\xnew)^{-1}
\end{bmatrix}
= \begin{bmatrix}
H(\xnew) & Q(\xnew) \\
Q(\xnew)^\top & V(\xnew)
\end{bmatrix} \\
\mbox{and} \quad
\dfrac{\partial\W_{n+M}}{\partial \xnewip} &
= \dfrac{\partial}{\partial \xnewip} \begin{bmatrix}
\Wn & w(\vecX_n, \xnew) \\
w(\vecX_n, \xnew)^{\top} & w(\xnew, \xnew)
\end{bmatrix}
= \begin{bmatrix}
\mathbf{0} & S(\xnew)\\
S(\xnew)^{\top} & T(\xnew)
\end{bmatrix}.
\end{align*}
Expressions for $H(\xnew)$, $Q(\xnew)$,  $V(\xnew)$, $S(\xnew)$ and
$T(\xnew)$, which are tedious,  are provided in Appendix \ref{sec:Igrad}. With
these quantities and Eq.~\eqref{equ:Inm}, the gradient of $I_{N+M}$ can be
expressed as:
\begin{align}
-\frac{\partial I_{N+M}}{\partial \xnewip} 
&=  \blu{\hat{\tau}^{2}} \left[ \tr(g(\xnew)\frac{\partial \Sigma(\xnew)}{\partial \xnewip} g(\xnew)^{\top}) + 2 \tr(Q(\xnew)\Sigma(\xnew) g(\xnew)^{\top}) \right. \nonumber \\
&\quad +  2\tr(Q(\xnew) w(\vecX_n, \xnew)^{\top}) + 2\tr(g(\xnew) S(\xnew)^{\top}) 
\label{equ:dI} \\
&\left. \quad - \tr(V(\xnew) w(\xnew, \xnew)) + 
		\tr(\Sigma(\xnew)^{-1} T(\xnew))\right]. \nonumber
\end{align}
\blu{Details for $\frac{\partial \Sigma(\xnew)}{\partial \xnewip}$ are provided in
Appendix \ref{sec:Igrad}.}

Finally, for Eq.~\eqref{equ:chainrule} we need  $\frac{ \partial \W_{n+M}}{\partial
\xnewip}$. Our earlier expression for $w(\mathbf{x}_i, \mathbf{x}_j)$ was generic, however
derivatives are required across each of $d$ input dimensions for the gradient so here
we acknowledge a separable kernel structure for completeness. Component
$\W_{n+M}^{(i,j)}$ follows
\[
w(\mathbf{x}_i, \mathbf{x}_j) 
= \int \limits_{\x \in D} c(\mathbf{x}_i, \mathbf{x}) c(\mathbf{x}_j, \mathbf{x})d\mathbf{x} 
= \prod_{k=1}^{d} \int \limits_{x \in [0,1]} c(\mathbf{x}_{i(k)}, x) c(\mathbf{x}_{j(k)}, x)dx 
= \prod_{k=1}^{d} w_k(\mathbf{x}_{i(k)}, \mathbf{x}_{j(k)}).
\] 
Appendix \ref{sec:Igrad} provides $w_k(\cdot,
\cdot)$ for a Gaussian kernel.  When differentiating with respect to
$\xnewip$, only the $(n+i)^\mathrm{th}$ row/column of $\frac{ \partial
\W_{n+M}}{\partial \xnewip}$ is non-zero. When $j \leq n$, those entries are
\[ 
\frac{\partial \W_{n+M}^{(n+i, j)}}{\partial \xnewip} 
= \frac{\partial w_p(\xnewip, \mathbf{x}_j)}{\partial \xnewip} 
\prod_{k=1, k\neq p}^{d} w_k(\vecxnew_{i(k)}, \mathbf{x}_{j(k)}).
\]
\blu{When $j >n$, swap $\x_j$ for $\vecxnew_{j-n}$}. A expression for
$\partial w_p$ is provided in the appendix.

\subsection{Implementation details and illustration}

Closed-form IMSPE and gradient in hand, selecting $M$-sized batches of new
runs becomes an optimization problem in $Md$ dimensions that can be off-loaded
to a library. When each dimension is constrained to $[0,1]$, i.e., assuming coded inputs,
we find that the L-BFGS-B algorithm \citep{BFGS} is appropriate, and generally
works well even in this high dimensional setting.  We use the
built-in {\tt optim} function in {\sf R}, taking care to avoid redundant
work in evaluating objective and gradient, which share structure.

\begin{figure}[ht!]
\centering
\includegraphics[scale=0.4,trim=0 15 0 20]{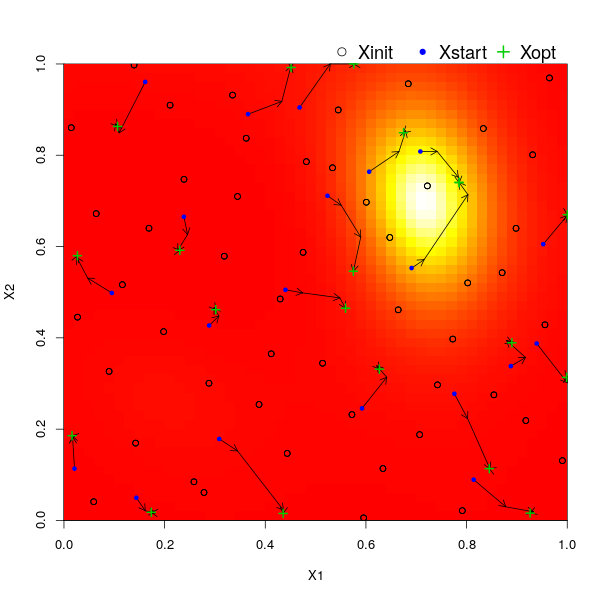}
\caption{Batch IMSPE optimization iterations from initial (blue dots)  to 
final (green crosses) locations.  Three optimization epochs are provided by
arrows. An overlayed heatmap shows the estimated standard deviation surface
$\sqrt{r(x)}$.}
\label{fig:IMSPE_opt}
\end{figure}

Figure \ref{fig:IMSPE_opt} provides an illustrative view of this new
capability \blu{by previewing the 2d toy example of Section \ref{sec:examples}}. We started with a space-filling design  $\vecXu$ in $[0,1]^2$, \blu{where 150 data points are evenly allocated on 50 unique design locations, shown as open circles}. The true noise surface, $r(x)$, was derived from a
standard bivariate Gaussian density with location $\mu = (0.7, 0.7)$ and scale
$\Sigma = 0.02 \cdot \mathbb{I}_2$.   The heatmap depicts a HetGP-estimated
standard deviation surface based on runs gathered at $\vecXu$.  Higher
noise regions are more yellow.  We then set out to calculate coordinates of a
new $M=20$ sized batch $\xnew$ via IMSPE. Search is initialized with a LHS,
shown in the figure as blue dots. Arrows originating from those dots show
progress of the derivative-based search broken into three epochs for dramatic
effect. Iterating to convergence requires hundreds of objective/gradient
evaluations in the $Md=40$-dimensional search space, but these each take a
fraction of a second because there are no large cubic operations.  At the
terminus of those arrows are green crosses, indicating the final locations of
the new batch $\xnew^\star$. Observe how some of these spread out relative to
one another and to the open circles (mostly in the red, low-noise region),
while others (especially near the yellow, high-noise region) are attracted to
each other.  At least one new replicate was found. Thus the  IMSPE criterion
strikes  a balance between filling the space and creating replicates, which
are good for separating signal from noise.

L-BFGS-B only guarantees a local minimum since the IMSPE objective is not a
convex function. Actually, IMSPE surfaces become highly multi-modal as more
points are added, with numbers of minima growing linearly in $n$, the number of
unique existing design elements, even in the $M=1$ case.  Larger batch sizes
$M > 1$ exacerbate this still further.  There is also a ``label-switching
problem''. (Swap two elements of the batch and the IMSPE is the same.) To
avoid seriously inferior local minima in our solutions for $\xnew^\star$ we
deploy \blu{a limited} multi-start scheme, starting multiple L-BFGS-B routines
simultaneously from novel sets of space-filling initial $\xnew^{(0)}$,
choosing the best at the end.

\section{Hunting for replicates}
\label{sec:merging}

Replication, meaning repeated simulations $Y(\mathbf{x})$ at fixed
$\mathbf{x}$, keeps cubic costs down [Eqs.~(\ref{equ:IN}) and
(\ref{equ:chainrule}), reducing from $N$ to $n$] and plays an integral role in
separating signal from noise \citep{Ankenman2010,hetGP1}, a win-win for
statistical and computational efficiency.  Intuitively, replicates become
desirable in otherwise poorly sampled high-variance regions \citep{HetGP2}.
Unfortunately, a numerical scheme for optimizing IMSPE will never precisely
yield replicates because tolerances on iterative convergence cannot be driven
identically to zero.  Consider again Figure \ref{fig:IMSPE_opt}, focusing now
on the two new design points in the yellow region which went to similar final
locations along their optimization paths. These look like potential
replicates, but their coordinates don't match.

One possible solution resolving near-replicates into actual ones is to
introduce a secondary set of tolerances in the input space, whereby closeness
implying ``effective replication'' can be deduced after the numerical solver
finishes.  This worked well for \citet{HetGP2}, in part because of an
additional \blu{replication-biased} lookahead device \citep{Ginsbourger2010} 
\blu{\citeauthor{HetGP2} used discrete search to check the degree to which future
replicates could reduce IMSPE.} But for us such tactics are unsatisfying on
several fronts: lookahead isn't manageable for $M \gg 1$ sized batches;
additional input tolerances are tantamount to imposing a grid; such a scheme
doesn't directly utilize IMSPE information; and finally whereas one-at-a-time
acquisition presents more opportunities to make adjustments in real-time, our
batch setting puts more eggs in one basket.  We therefore propose the
following post-processing scheme on each batch which we call ``backtracking''.

\subsection{Backtracking via merge}

For a new batch of size $M$, the possible number of new replicates ranges from
zero to $M$.  L-BFGS-B optimization yields $M$ unique coordinate tuples, but some
may be very close to one another or the $n$ existing unique sites. Below we
verbalize a simple greedy scheme for ordering and valuing those $M$ locations
as potential ``effective replicates''.  Choosing among those alternatives
happens in a second phase, described momentarily in Section
\ref{sec:selecting}.

Begin by recording the IMSPE of the solution $\xnew_M \equiv \xnew^\star$
provided by the optimizer: $I_{n+M}(\xnew_M)$.  This corresponds to the
no-backtrack/no-replicate option.  Set iterator $s=0$ so that $\xnew_{m_s}$
refers to this potential batch with $m_s = M$ unique design elements and let
$d_s = 0$.  Move to the first iteration, $s=1$.  Among the $m_{s-1}$ unique
sites in $\xnew_{m_{s-1}}$, find the one which has the smallest minimum
distance $d_s$ to other unique elements in $\xnew_{m_{s-1}}$ and existing
sites $\vecXu$, with ties broken arbitrarily.  Entertain a new batch
$\xnew_{m_s}$ by merging sights involved in that minimum $d_s$-distance pair. 
If both are a member of the new batch $\xnew_{m_{s-1}}$, then choose a
midway value for their new setting(s) in $\xnew_{m_s}$.  Otherwise, take the
location from the existing (immovable) unique design element from
$\vecXu$. Both imply $m_s = M - s$.  Calculate $I_{n+m_s}(\xnew_{m_s})$. 
Increment $s \leftarrow s + 1$ and repeat unless $s=M$. 


\begin{figure}[ht!]
	\centering
	\includegraphics[scale=0.46,trim=0 50 0 20]{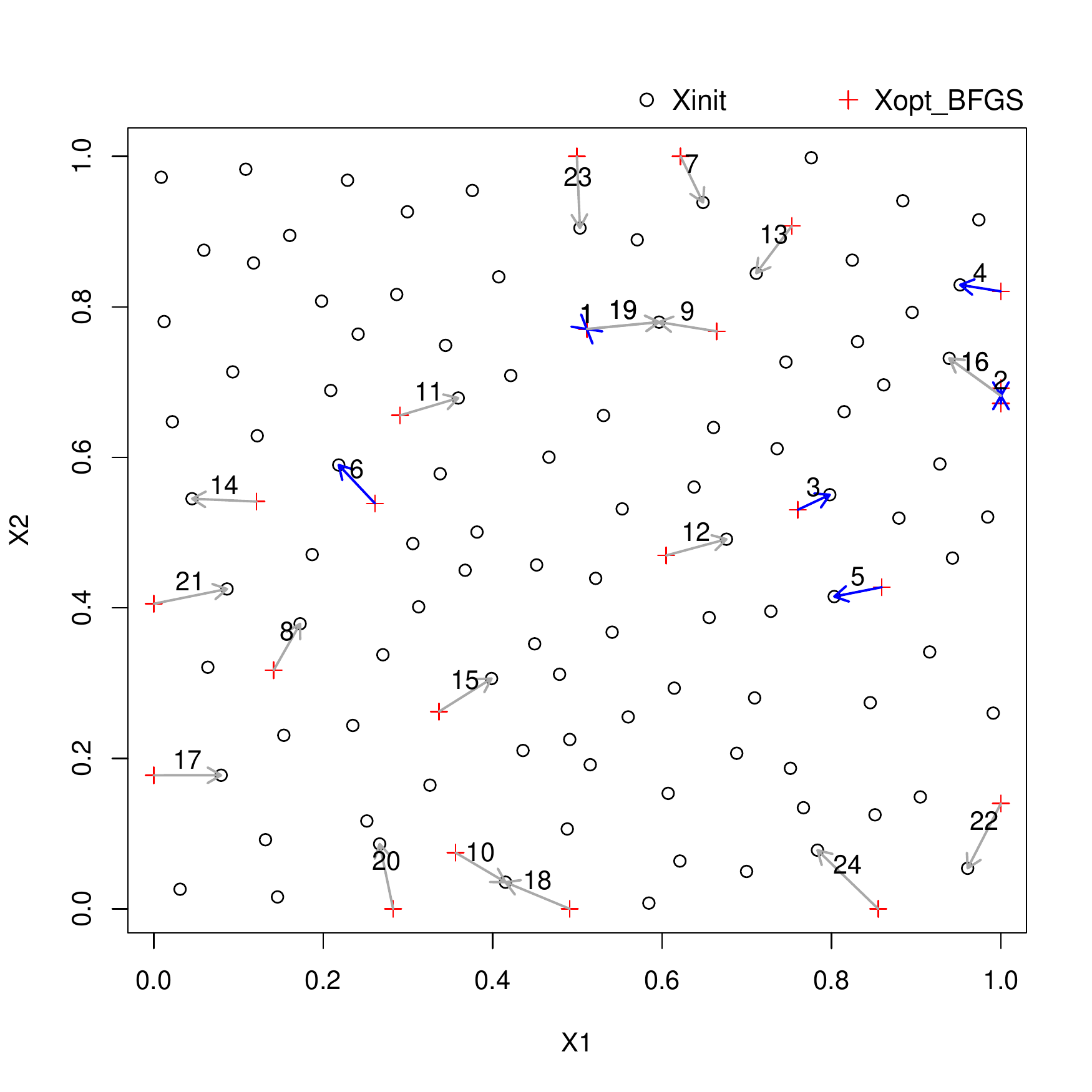}
	\includegraphics[scale=0.46,trim=15 50 0 20]{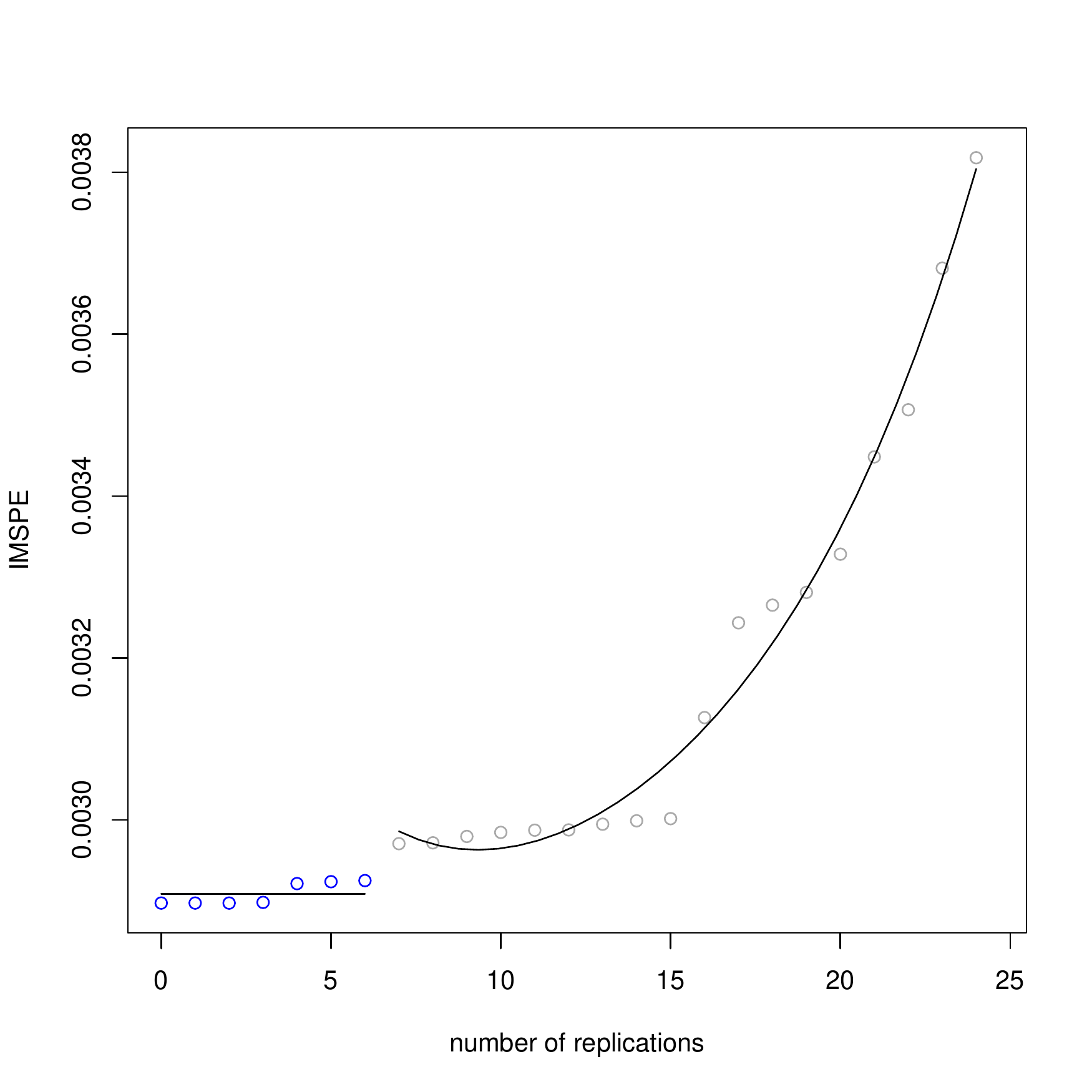}
	\caption{{\em Left:} backtracking with merge; gray arrows connect optimal $\xnew_{m_s}$ with numbers indicating $s=1,\dots, M$; Right: IMSPE changes over numbers of replicates. Merging steps that are finally taken are shown in blue. Fitted segmented regression lines are overlaid.}
	\label{fig:merging}
\end{figure}

Figure \ref{fig:merging} provides an illustration; settings of $f(\mathbf{x})$
and $r(\mathbf{x})$ mirror Figure \ref{fig:IMSPE_opt}.   The existing design
$\vecXu$ has $n = 100$ unique elements, shown as open circles in the left
panel.  Each run is replicated three times so that $N=300$.   A new batch of
size $M = 24$ is sought. Red crosses represent optimized $\xnew_{m_0} =
\xnew_M$ from L-BFGS-B. Numbered arrows mark each backtracking step.  Observe
that the first two of these (almost on top of one another near the right-hand
boundary) involve novel batch elements, whereas all others involve one of the
$n$ existing sites.  Aesthetically, the first five or so look reasonable,
being nearby the high variance (top-right) region. Replication is essential in
high-variance settings.

\subsection{Selecting among backtracked batches}
\label{sec:selecting}

To quantify and ultimately automate that eyeball judgment, we investigated
$I_{n+m_s}(\xnew_{m_s})$ versus $s$, the number of replicates in the new
batch. The right panel of Figure \ref{fig:merging} shows the pattern
corresponding to the backtracking steps on the left. Here, the sequence of
$I_{n+m_s}(\xnew_{m_s})$ values is mostly flat for $s = 0, \dots, 3$, then
increasing thereafter. We wish to minimize IMSPE, except perhaps preferring
exact replicates when IMSPEs may technically differ but are very similar. 
Aesthetically, that ``change point'' happens at $s=7$ where IMSPE jumps into a
new and higher regime.

To operationalize that observation we experimented with a number of change
point detection schemes.  For example, we tried the {\tt tgp} \citep{tgp,tgp2}
family of Bayesian treed constant, linear, and GP models.  This worked great,
but was overkill computationally.  We also considered placing $d_s$, the
minimizing backtracked pairwise distances, on the $x$-axis rather than
$s$-values.  Although the behavior with this choice was distinct, it yielded
more-or-less equivalent selection on broad terms.

We ultimately settled on the following custom scheme recognizing that the
left-hand regime was usually constant (i.e., almost flat), and the right-hand
regime was generally increasing.\footnote{BFGS is a local solver and
backtracking is greedy, both contributing to potential for non-monotonicity.}
To find the point of shift between those two regimes,  we fit $M+1$
two-segment polynomial regression models, with break points $s=0, \dots, M$
respectively, with the first regime (left) being of order zero (constant) and
the second (right) being of order four.  We then chose as the location
$\hat{s}$ the one whose two fits provide lowest in-sample MSE. The optimal
pair of polynomial fit pairs are overlaid on the right panel of Figure
\ref{fig:merging}, with groups color-coded to match arrows in the left panel.
 
 \begin{figure}[ht!]
\centering
\includegraphics[scale=0.45,trim=0 15 10 40,clip=TRUE]{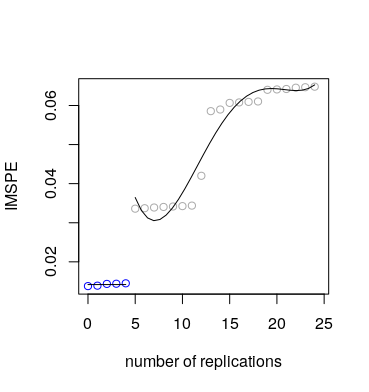}
\includegraphics[scale=0.45,trim=30 15 10 40,clip=TRUE]{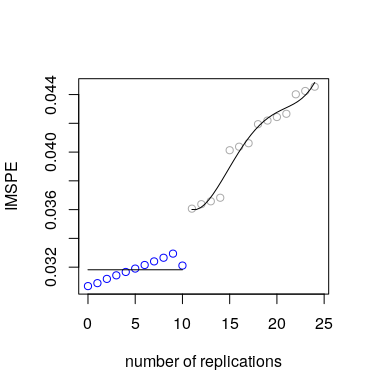}
\includegraphics[scale=0.45,trim=30 15 10 40,clip=TRUE]{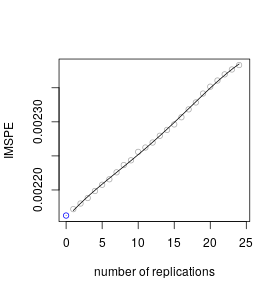}
\includegraphics[scale=0.45,trim=30 15 10 40,clip=TRUE]{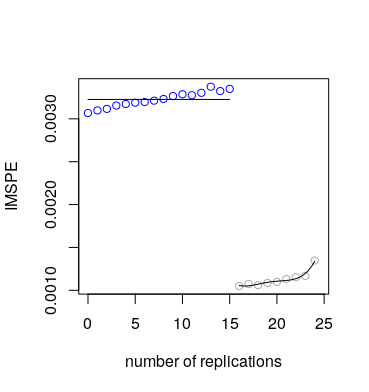}
\caption{Three selected scatter plots of IMSPE versus number of replicates with best change-point fitted regression lines overlaid. Colors match arrows in Figure \ref{fig:merging}.}
\label{fig:split}
\end{figure}

Figure \ref{fig:split} shows four other examples under the same broad settings
but different random initial $n$-sized designs.  The situation in the left
panel matches that of Figure \ref{fig:merging} and is by far the most common.
The 2$^\mathrm{nd}$ panel depicts a setting where zero replicates is best but
the two-regression scheme nevertheless identifies a midway change-point
suggesting a bias toward finding at least some replicates. The 3$^\mathrm{rd}$
panel shows the case where no replicates are included. The right panel
indicates an \blu{uncommon}, opposite extreme.  Note the small range of the
IMSPE axis ($y$-axis). When the right-hand regime has uniformly lower IMSPE
than the left-hand one, \blu{as may happen if merging identifies another local
optima}, we take $\hat{s}$ as the choice minimizing IMSPE in the right-hand
regime.

\section{Benchmarking examples}
\label{sec:examples}
	
Here we illustrate and evaluate our method on an array of test problems.  We
have four examples total.  Two are relegated to Appendix \ref{app:examples}:
one mirroring the 1d example from \citet{HetGP2}; another involves a 4d ocean
simulator from \citet{ocean}.  The other two, showcased here, include a 2d toy
problem and an 8d ``real simulator'' from inventory management.  Metrics
include out-of-sample root mean-squared prediction error (RMSPE), i.e.,
matching our IMSPE acquisition heuristic, and a proper scoring rule
\citep[][Eq.~(27)]{score} combining mean and uncertainty quantification
accuracy, which for GPs reduces to predictive log likelihood. We also consider
computing time and number of unique design elements, $n$, over total
acquisitions $N$.  Our gold standard benchmark is the ``pure sequential''
($M=1$) adaptive lookahead scheme of \citeauthor{HetGP2}, however when
relevant we also showcase other special cases.  Our goal is not to beat that
benchmark.  Rather we aim to be competitive while entertaining $M=24$-sized
batches, representing the number of cores on a single supercomputing node. 

\subsection{2d toy example} 

\begin{figure}[!ht]
\centering
\includegraphics[trim={10 20 0 40},width=0.43\textwidth]{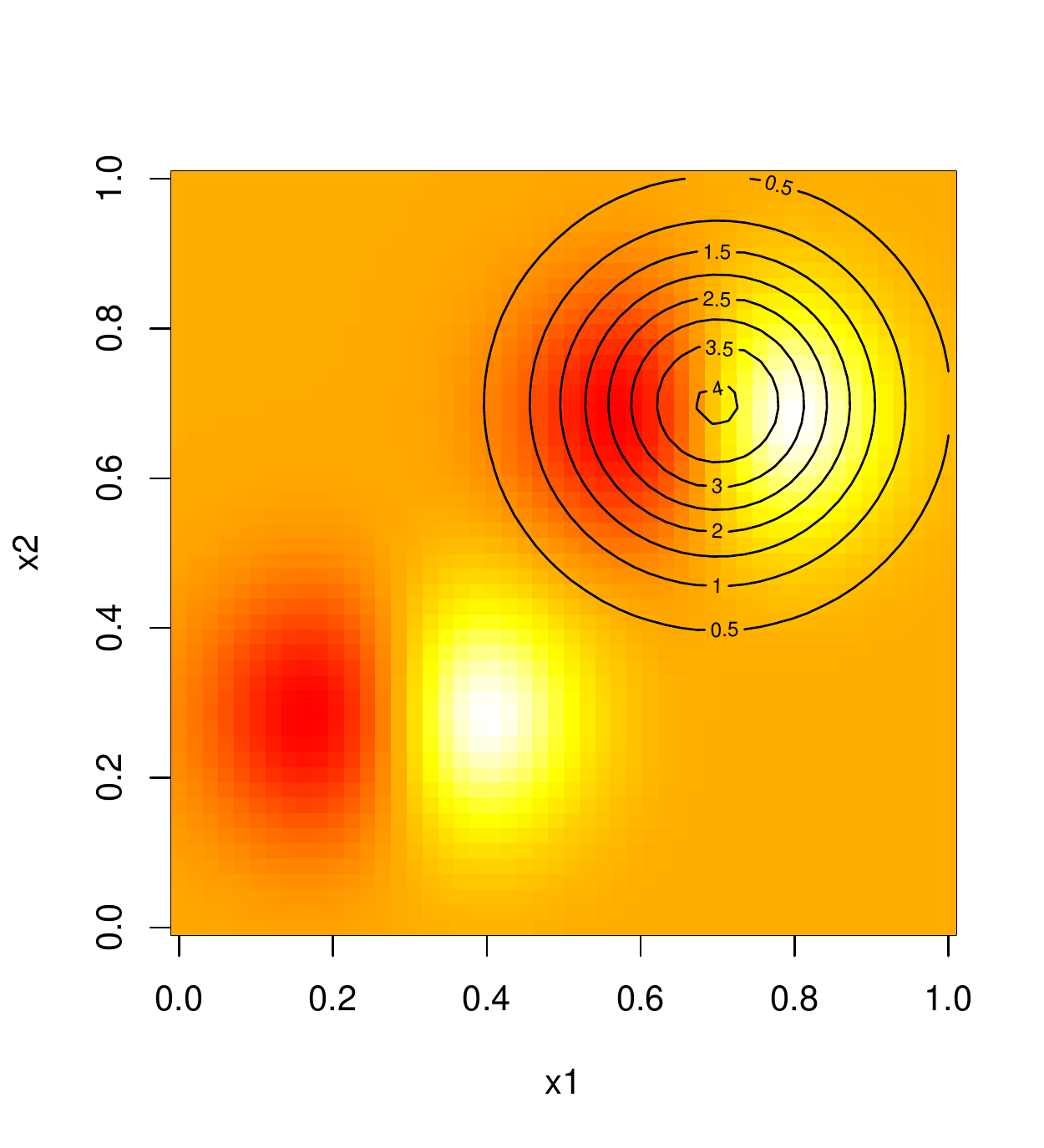}
\caption{The heatmap shows the mean surface $f(\x)$. Lighter colors correspond
to higher values. Contours of $r(\x)$ are overlaid. }
\label{fig:2dtruth}
\end{figure}

Elements of this example have been in play in previous illustrations,
including Figures \ref{fig:IMSPE_opt}--\ref{fig:merging}. The true mean
function $f(\x)$ is defined as:
\[
f(\mathbf{x}) = f(x_1, x_2) = 20 \left[ \frac{a_1}{\exp(a_1^2 + a_2^2)} 
+ \frac{a_3}{\exp(a_3^2 + a_4^2)} \right],
\]
where $ a_1 =  6 x_1 - 4.1$, $a_2 =  6 x_2 - 4.1$, $a_3 =  6 x_1 - 1.7$, and
$a_4 =  6 x_2 - 1.7$. The true noise surface, $r(\x)$, is a bivariate Gaussian
density with location $\mu = (0.7, 0.7)$ and scale $\Sigma = 0.02 \cdot
\mathbb{I}_2$. Figure \ref{fig:2dtruth} provides a visual using color for
$f(\x)$ and contours for $r(\x)$.  We deliberately made the mean surface have
the same signal structure at the bottom left and top right regions. However,
the top right region is exposed to high noise intensity while the bottom
left region is almost noise-free, creating distinct signal-to-noise regimes.

Design aspects of our experiment(s) were set up as follows.  We begin with an
$n_0=20$-sized maximin--LHS with five replicates upon each for $N_0=100$ total
simulations.  This is followed by ten batches of
IMSPE-acquisition with backtracking for 240 new runs
($N=340$ total).  
\begin{figure}[ht!]
\centering
\includegraphics[width=\textwidth,trim=0 25 0 5]{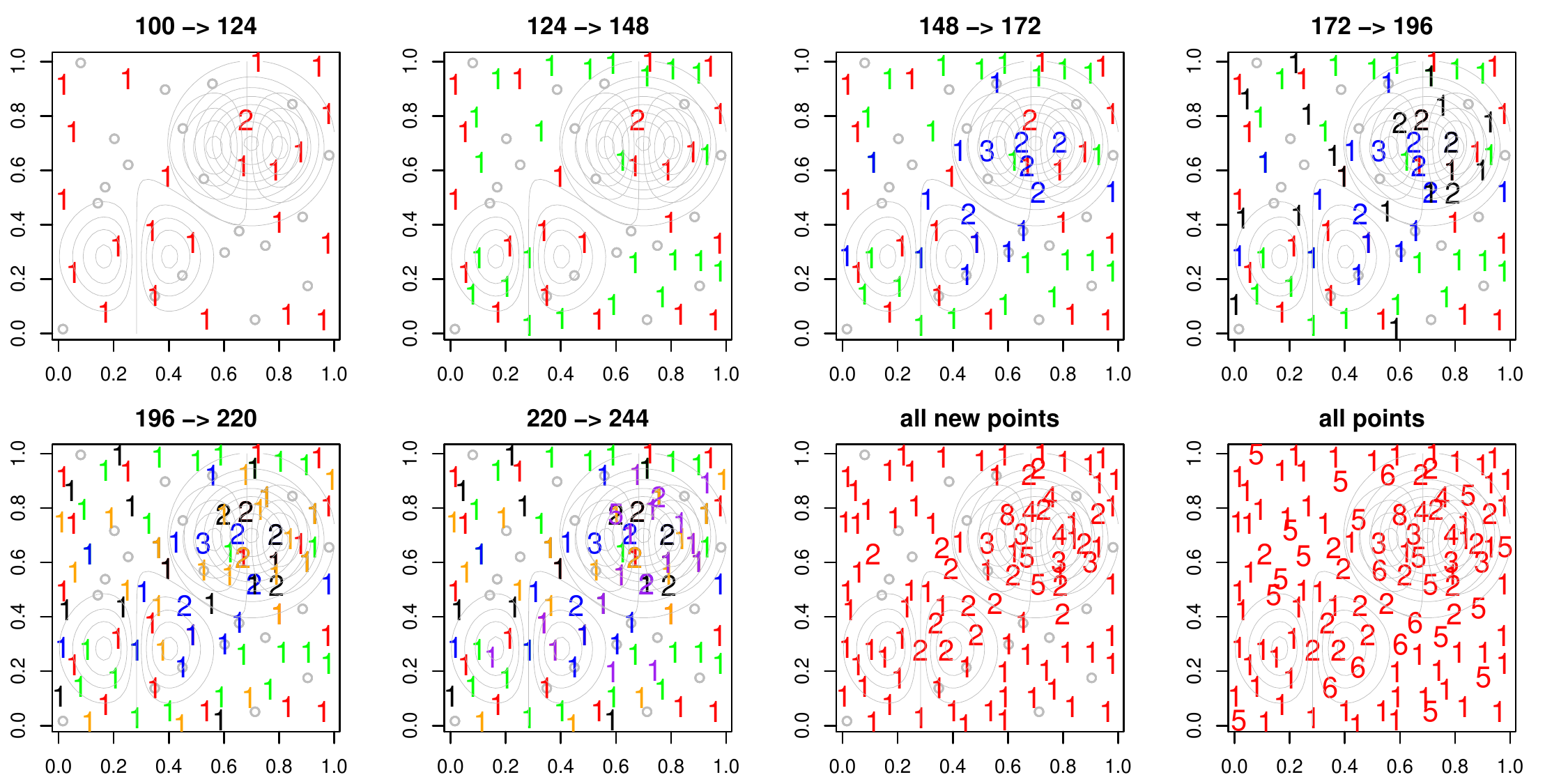}
\caption{IMSPE design in batches: gray dots are initial design points; gray contours show signal and noise contrast; numbers indicate replicate multiplicity. The last two panels summarize all new points from 6 batches and all design points respectively.}
\label{fig:2dresult}
\end{figure}
Figure \ref{fig:2dresult} shows how the first six batches
distributed in the input space, with one panel for each.   Color is used to
track batches over accumulated runs; numbers indicate degrees of replication.  For
example, the first batch had two replicates (one at a unique input, one at an
existing open circle), whereas the third batch had many more.  Observe that as
batches progress, more replicates and more unique locations cluster near the
noisy top-right region of the input space.  The final two panels summarize all
(new) points involved in those first six batches, including the initial
design.
	
\begin{figure}[ht!]
\centering
\includegraphics[width=0.99\textwidth]{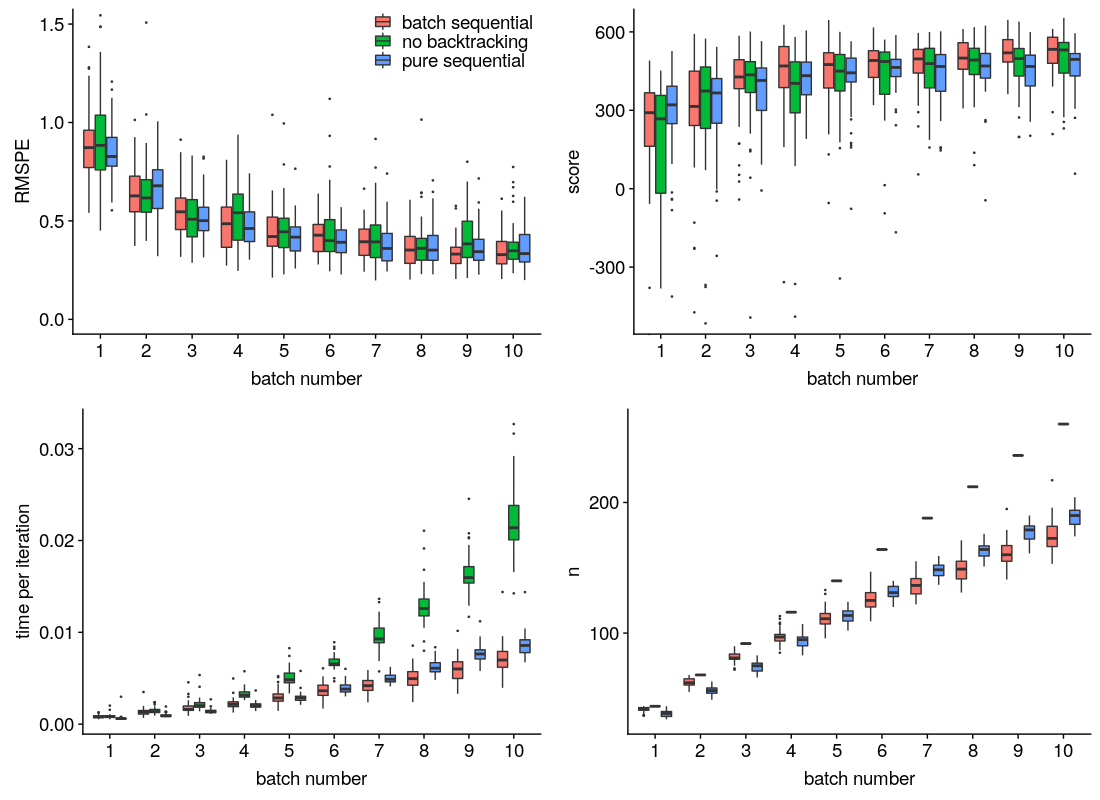}
\caption{Results of RMSPE, score, time per iteration in fitting hetGP model,
and the aggregate number of unique design locations from 50 MC repetitions.}
\label{fig:2dcp} 
\end{figure}

Figure \ref{fig:2dcp} offers a comparison to \cite{HetGP2}'s pure sequential
($M=1$) strategy in a fifty-repetition MC exercise. Randomization is over the
initial maximin--LHS, noise deviates in simulating the response, and novel LHS
testing designs of size $N=500$.  We also include a ``no backtracking''
comparator, omitting the search for replicates step(s) described in Section
\ref{sec:merging}.   For the pure sequential benchmark, we calculate RMSPE and
score after every 24 subsequent steps to make it comparable to our batches. In
terms of RMSPE, all three methods perform about the same.  Under the other
three metrics, batch-with-backtracking is consistently better than the
non-backtracking version: more replicates, faster HetGP fits due to smaller
$n$, and higher score after batch three. The degree of replication yielded by
backtracking is even greater than the pure sequential scheme after batch four.
Also from batch four, batch IMSPE outperforms pure sequential design on score.

\subsection{Assemble-to-order}
	
The assemble-to-order (ATO) problem \citep{Hong:2006} involves a queuing
simulation targeting inventory management scenarios. It was designed to help
determine optimal inventory levels for eight different items to maximize
profit.  Here we simply treat it as blackbox response surface. Although the
signal-to-noise ratio is relatively high, ATO simulations are known to be
heteroskedastic \citep{hetGP1}. We utilized the {\tt MATLAB} implementation
described by \citet{Xie:2012} through {\tt R.matlab} \citep{R.matlab} in {\sf
R}. Our setup duplicates the MC of \citet{HetGP2} in thirty
replicates, in particular by initializing with a $n_0=100$-sized random design
in the 8d input space, paired with random degrees of replication $a_i \sim
\mathrm{Unif}\{1,\dots,10\}$ so that the initial design comprised about $N_0
\approx 500$ runs. \citeauthor{HetGP2} then performed about 1500 acquisitions
to end at $N=2000$ total runs.  We performed sixty-three $M=24$-sized batches
to obtain about 2012 runs.
	
Since the 8d inventory input vector must be comprised of integers
$\{0,\dots,20\}$, we slightly modified our method in a manner similar to
\citeauthor{HetGP2}: inputs are coded to $[0,1]$ so that IMSPE optimization
transpires in an $M \times [0,1]^8$ space. When backtracking, merged IMSPEs
are calculated via rounded $\xnew_{m_s}^{\mathrm{int}}$ on the natural scale.

\begin{figure}[!ht]
\centering
\includegraphics[scale=0.47,trim=15 10 15 15]{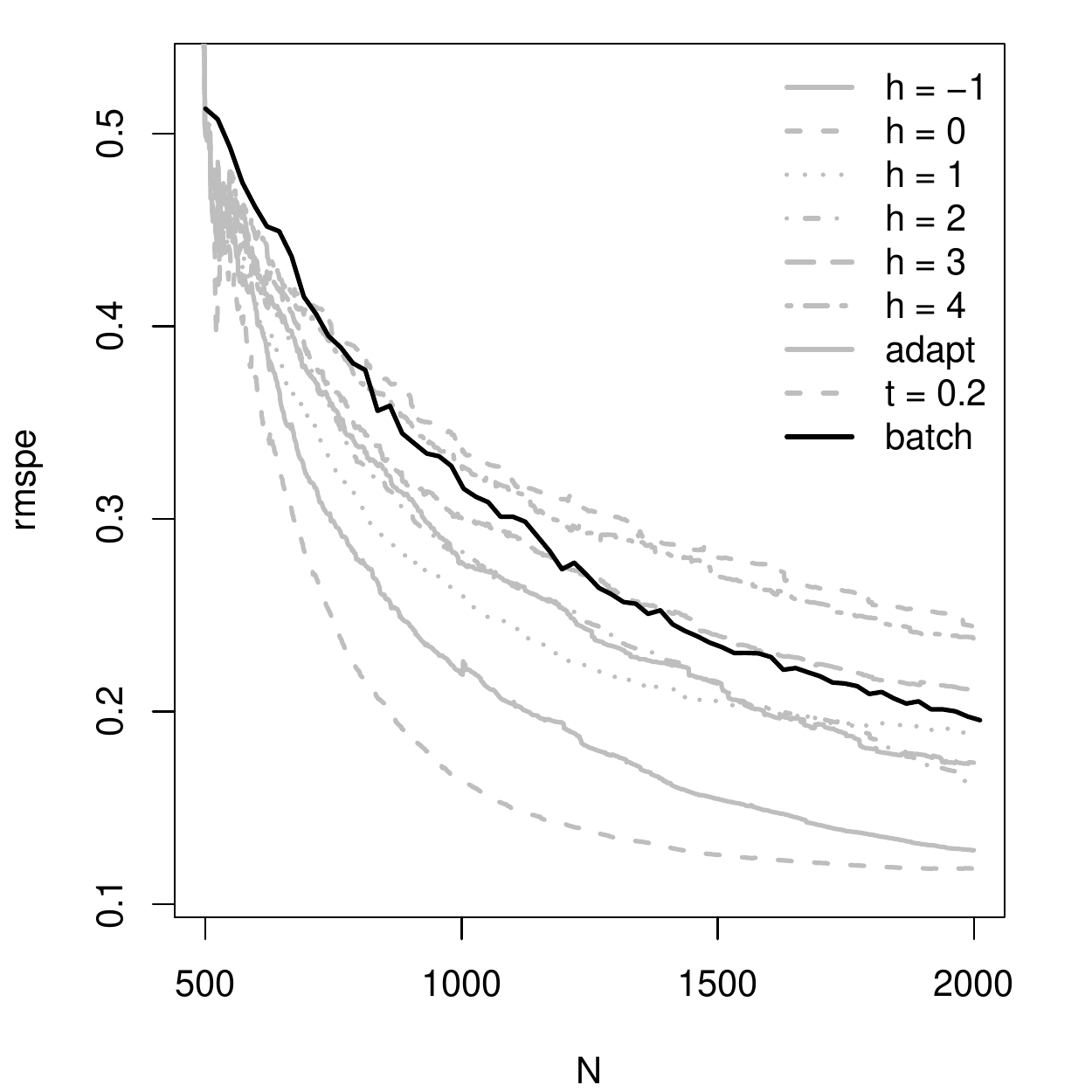}
\includegraphics[scale=0.47,trim=20 10 15 15]{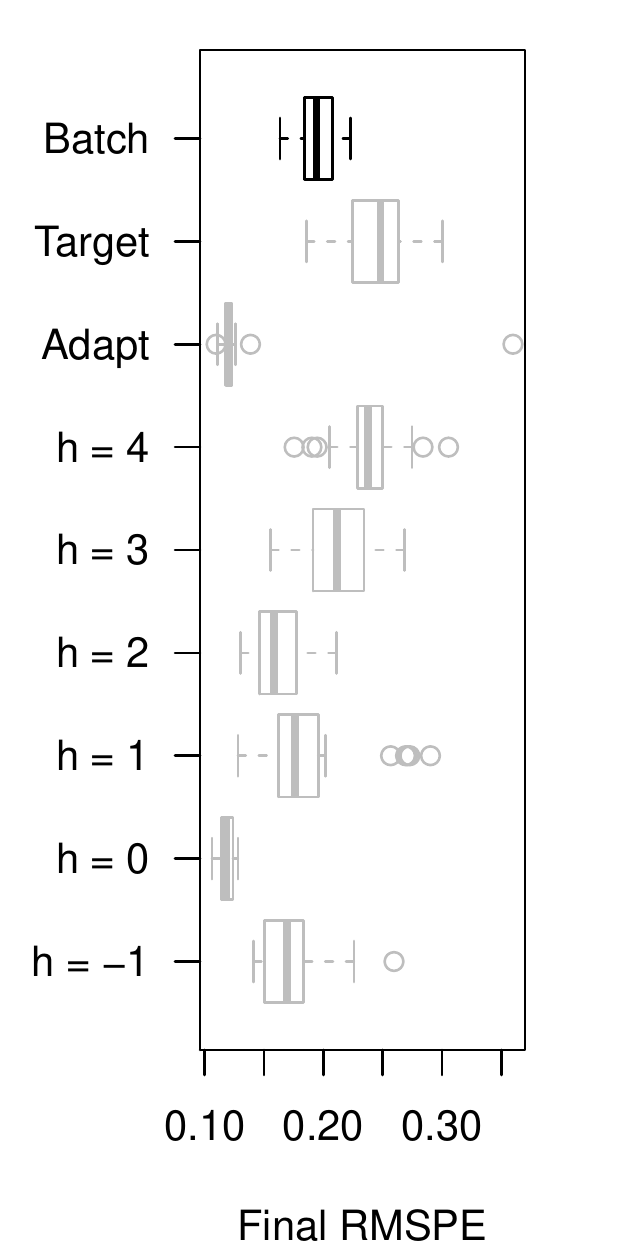}
\includegraphics[scale=0.47,trim=10 10 15 15]{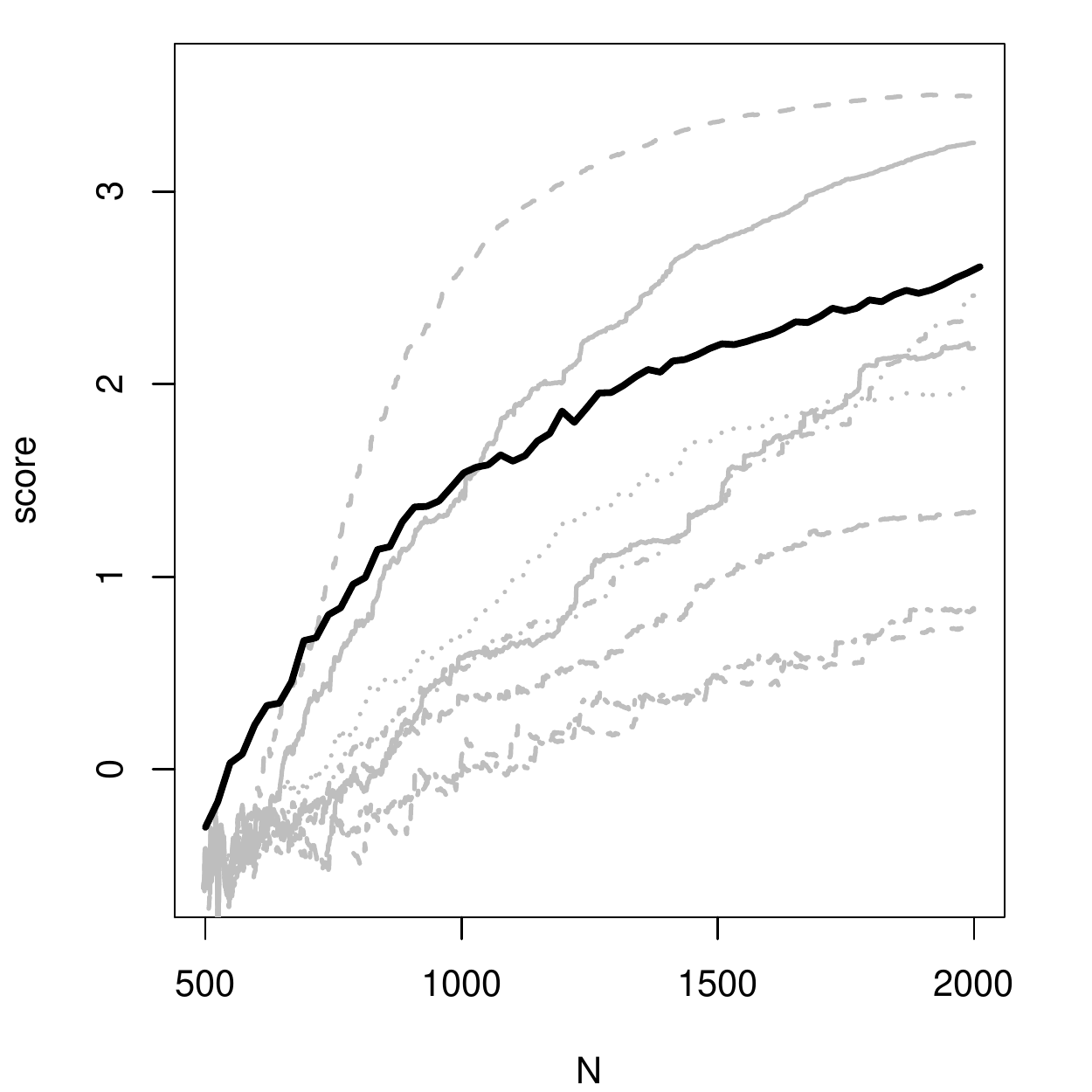}
\includegraphics[scale=0.47,trim=20 10 20 15]{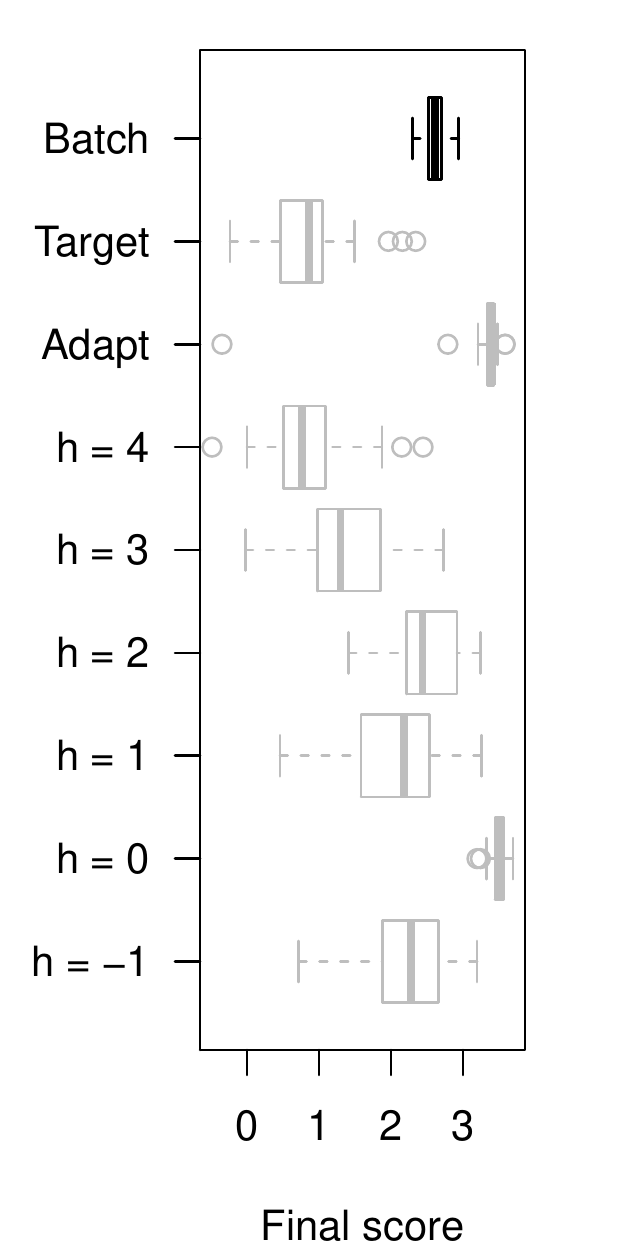}
\caption{RMSPE and score over design size $N$ from 30 MC repetitions.}
\label{fig:ato}
	\end{figure}

Figure \ref{fig:ato} shows progress in terms of average RMSPE and score
mimicking the format of the presentation of \citeauthor{HetGP2}, whose
comparators are duplicated in gray in our updated version.  There are eight
gray variations, representing multiple lookahead horizons ($h$) and two
automated horizon alternatives, with ``Adapt'' being the gold standard.  In
terms of RMSPE, our batch method makes progress more slowly at first, but
ultimately ends in the middle of the pack of these pure sequential
alternatives.  In terms of score, we start out the best, but end in the third
position. Apparently, our batch scheme is less aggressive on reducing
out-of-sample mean-squared error, but better at accurately assessing
uncertainty.  In the 30 MC instances our average number of {\em new}
replicates per unique site was 1.64 (min 0, max 5), leading to a mean of
$n=1610$ (min 1606, max 1612). This is a little higher (lower replication)
than $n=1086$ (min 465, max 1211) reported by \citeauthor{HetGP2} for
``Adapt''.  Again, we conclude that our batch method is competitive despite
being faced with many fewer opportunities to re-tune the strategy over
acquisition iterations.

\section{Delta smelt}
\label{sec:deltasmelt}

Encouraged by these results, and by simulations in Section \ref{sec:examples},
we return now our motivating delta smelt ABM application.  Time and allocation
limits meant only one crack at this, so we did one last ``sanity check'',
extending the pilot study with batch acquisitions, before embarking on a big
batch-sequential simulation campaign.  This analysis, which was also
encouraging, is described in detail in Appendix \ref{app:dist}.

\subsection{Setup and acquisitions}

For our``full'' simulation campaign and analysis of the delta smelt ABM where
we explored a ten-dimensional input space on a 7d manifold.  See Table
\ref{tab:smeltfullstudy}, augmenting  Table \ref{tab:deltasmelt} with a new
column. This analysis expanded the effective input domain by three, and
involved slightly adjusted ranges and relationships between the original
inputs.  Specifically, we extended $m_y$ and began to vary $P_{l,2}$,
$P_{a,3}$, and $P_{a,4}$ with dependencies $P_{p,6}=P_{p,2}\times 1.75 +
0.05$, $P_{j,3}=P_{j,6}$, and $P_{a,3}=P_{a,4}$. Inputs $m_l$, $m_p$, and
$m_a$ remain fixed at their default values.

\begin{table}[h!]
	\centering
	\begin{tabular}{l c c c c c}
		\hline
		symbol & range   &default &pilot study &full study\\
		\hline \hline
		$m_y$  
		&[0.01,0.50]  &0.035 &0.035 & [0.02, 0.05]\\
		$m_l$     &[0.01, 0.08]   &0.050 &0.050 & 0.050 \\
		$m_p$ &  [0.005, 0.05]	 &0.030 &0.030 & 0.030\\
		$m_j$ & [0.001, 0.025] &0.015 &[0.005,0.030] & [0.005, 0.030]\\
		$m_a$ & [0.001, 0.01]  &0.006 &0.006 & 0.006\\
		\hline
		$m_r$ & [0.005, 0.05]	&0.020 &[0, 0.05] & [0, 0.1]\\
		\hline
		$P_{l,2}$ & [0.10, 20.0]	   &0.200 &0.200 & [0.1, 0.5]\\
		$P_{p,2}$ & [0.10, 20.0]      &0.800 &[0.10, 1.84] &[0.10, 1.84]\\ 
		$P_{p,6}$ & [0.10, 20.0]	   &1.500 & $P_{p,2}$ & 1.75$P_{p,2}  + 0.05$\\
		$P_{j,3}$ & [0.10, 20.0]	  &0.600 &[0.1, 1.5] & [0.1, 1.5]\\ 
		$P_{j,6}$ &  [0.10, 20.0]	   &0.600 &  $P_{j,3}$ &  $P_{j,3}$\\
		$P_{a,3}$ & [0.01, 20.0]	   &0.070 &0.070 & [0.05, 0.15]\\
		$P_{a,4}$ &  [0.01, 5.0] 	   &0.070 &0.070 &  $P_{a,3}$\\
		\hline	
	\end{tabular}
	\caption{Augmenting Table \ref{tab:deltasmelt} to show the parameter
		settings of the ``full'' experiment.}
	\label{tab:smeltfullstudy}
\end{table}

To explore the 7d input space, we begin with maximin-LHS of size $n_0=192$,
each with five replicates for a total of $N_0 = 960$ initial runs. We aim to
more than double this simulation effort, collecting a total of $N= 2016$ runs,
by adding 44 subsequent batches of size $M=24$.  This took a total of 50 days,
requiring slightly more than one day per batch, including HetGP updates, IMSPE
evaluation and backtracking, and any time spent waiting in the queue on the
ARC HPC facility at Virginia Tech.  Inevitably, some hiccups prevented a fully
autonomous scheme. In at least one case, what seemed to be a conservative
request of 10 hours of job time per batch (of runs that usually take 4-6
hours) was insufficient.  We had to manually re-run those failed simulations,
and subsequently upped requests to 14 hours.  This bigger demand led to longer
queuing times even though the average execute time was at par with previous
campaigns.

When training the HetGP surrogate we used responses $y_i = \log \lambda_i$ for
nonzero simulation outputs.  Any zeros are replaced with $y_i = \log
\frac{1}{2}\min_{\{i:\lambda_i > 0\}} \lambda_i$ where $\{i:\lambda_i > 0\}$
represents the subset of $\{1,\dots,N\}$ indexing positive outputs. This lead
to slightly different $y$-axis scales for visuals compared to
Section \ref{sec:pilot}. An adaptive scheme for handling zeros was
necessitated by the dynamic nature of the arrival of $\lambda$-values
furnished over the batches of sequential acquisition -- in particular of ones
smaller than those obtained in the pilot study.

\begin{figure}[ht!]
	\centering
	\small
	\tabcolsep=0.11cm
	\begin{tabular}{ccc}
		\;\;\; mean surface & \;\; variance surface & \;\; 1d predictive bands \\
		\includegraphics[trim={10 20 0 25},width=52mm]{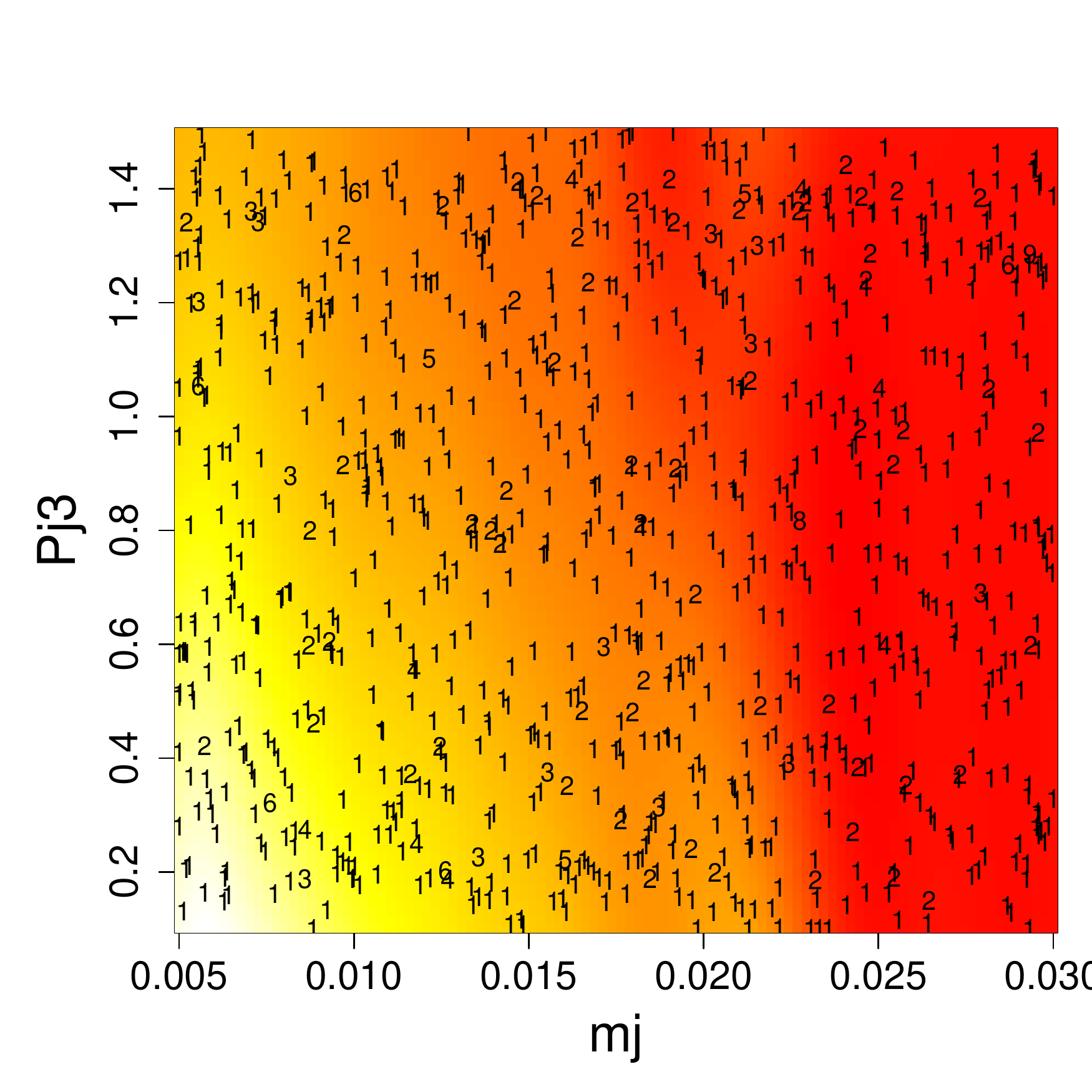} & 
		\includegraphics[trim={10 20 0 25},width=52mm]{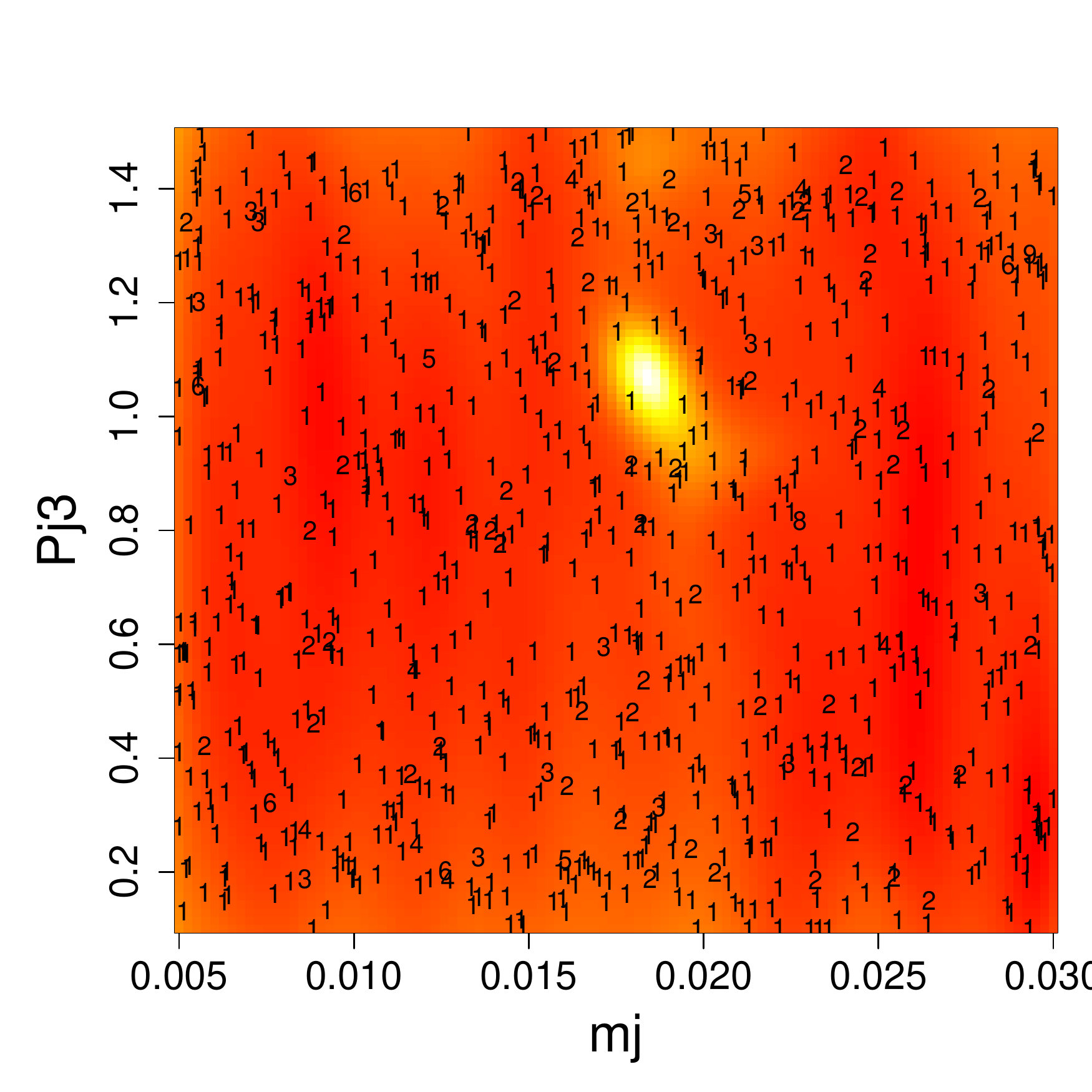} &  
		\includegraphics[trim={10 20 0 25},width=52mm]{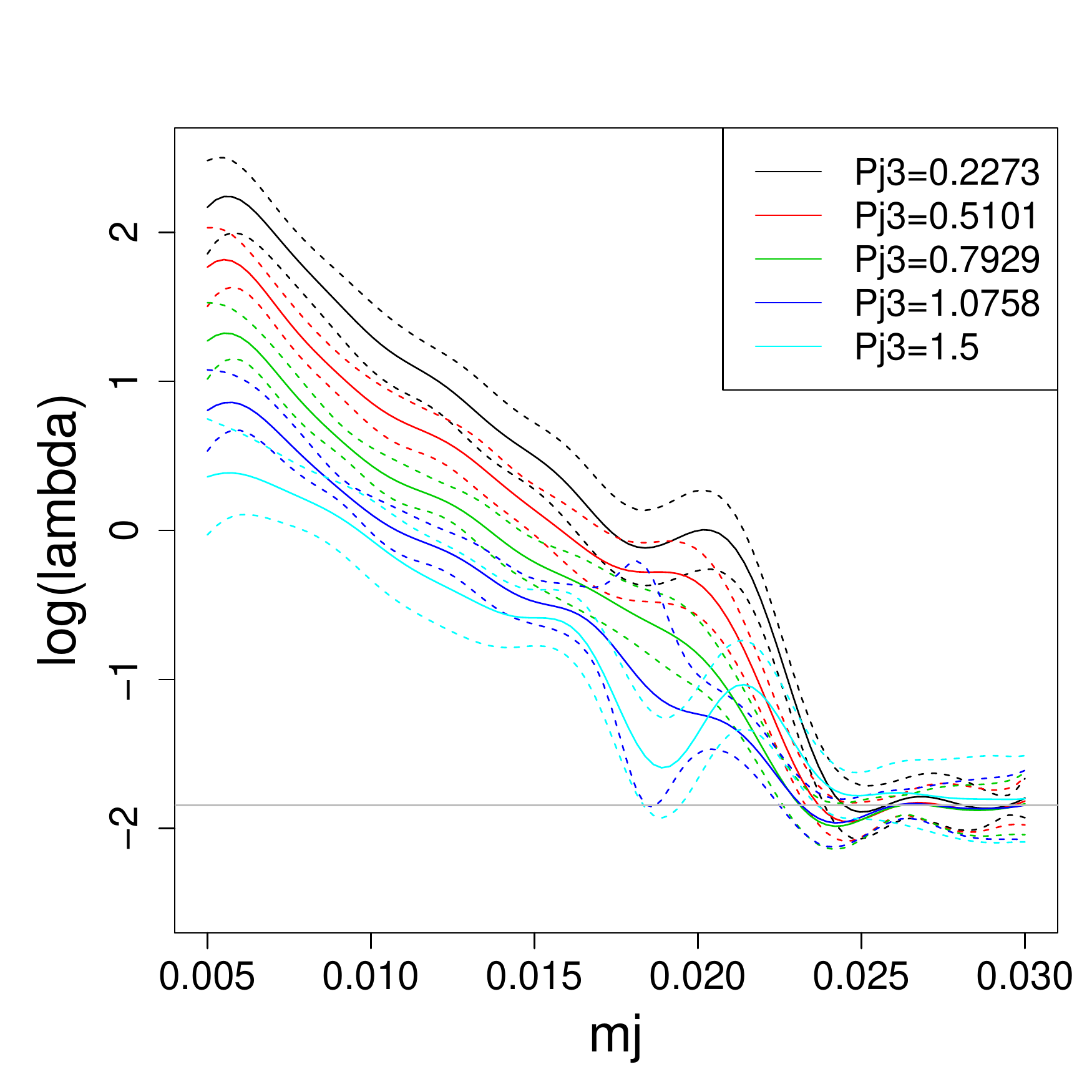}\\

		\includegraphics[trim={10 20 0 25},width=52mm]{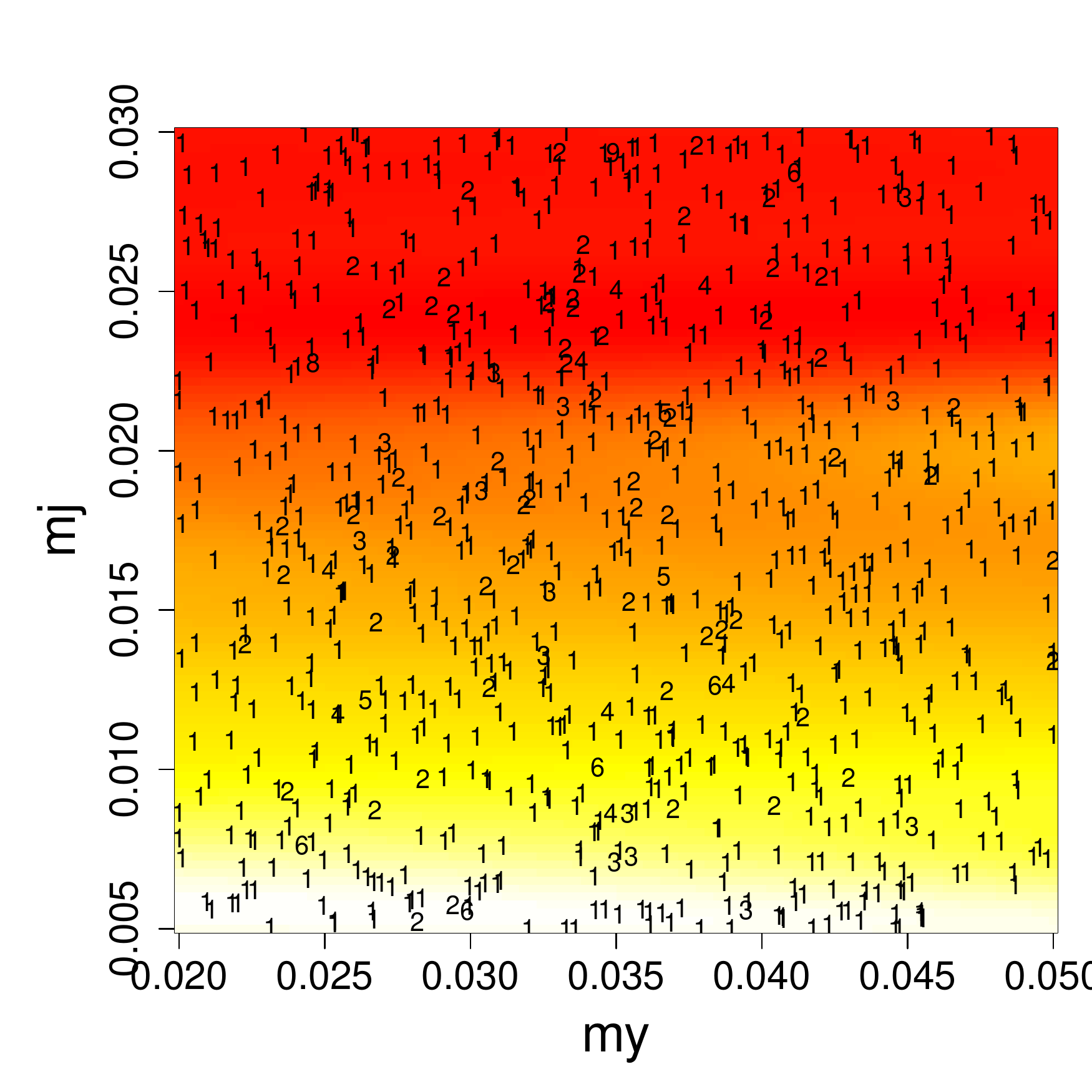} & 
		\includegraphics[trim={10 20 0 25},width=52mm]{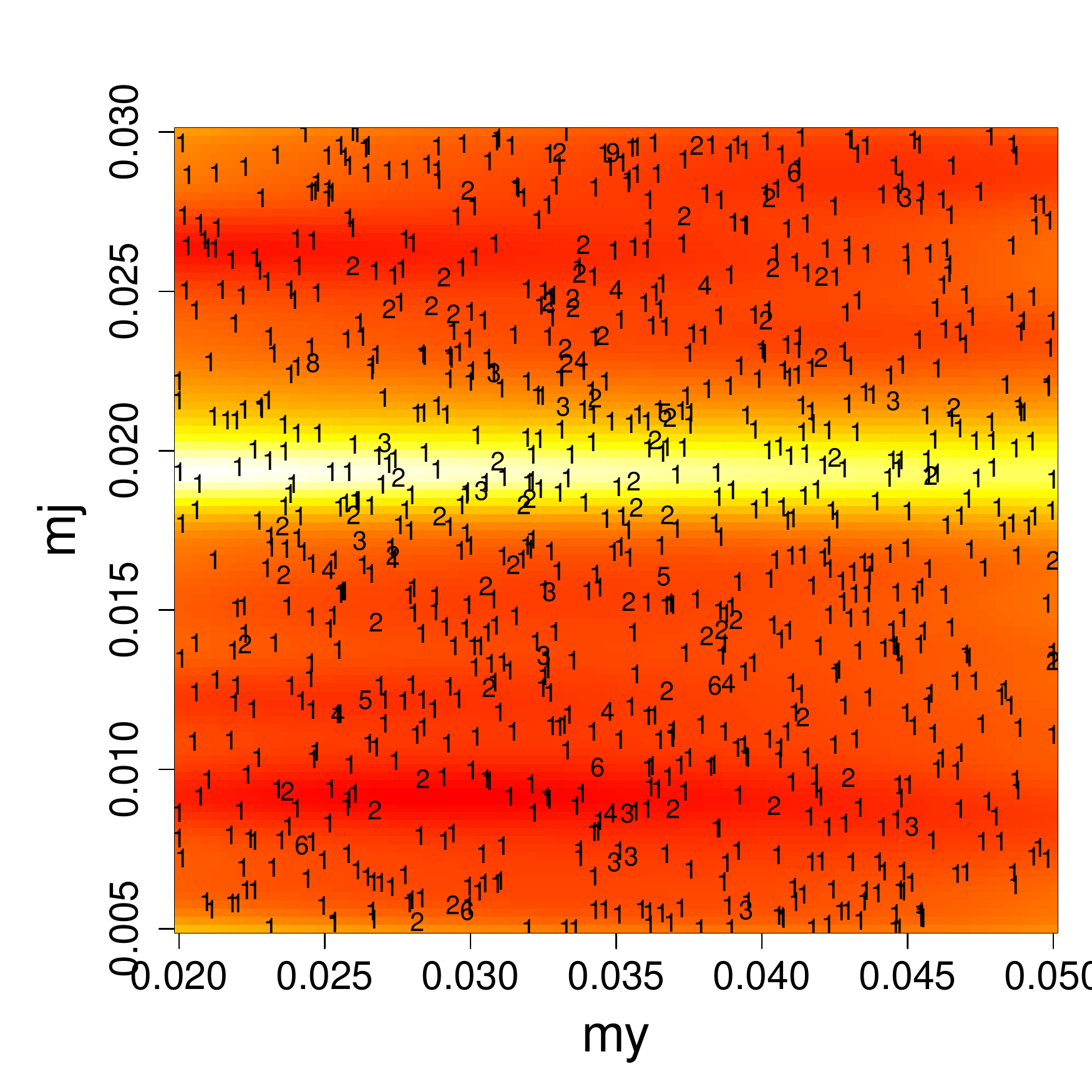} &  
		\includegraphics[trim={10 20 0 25},width=52mm]{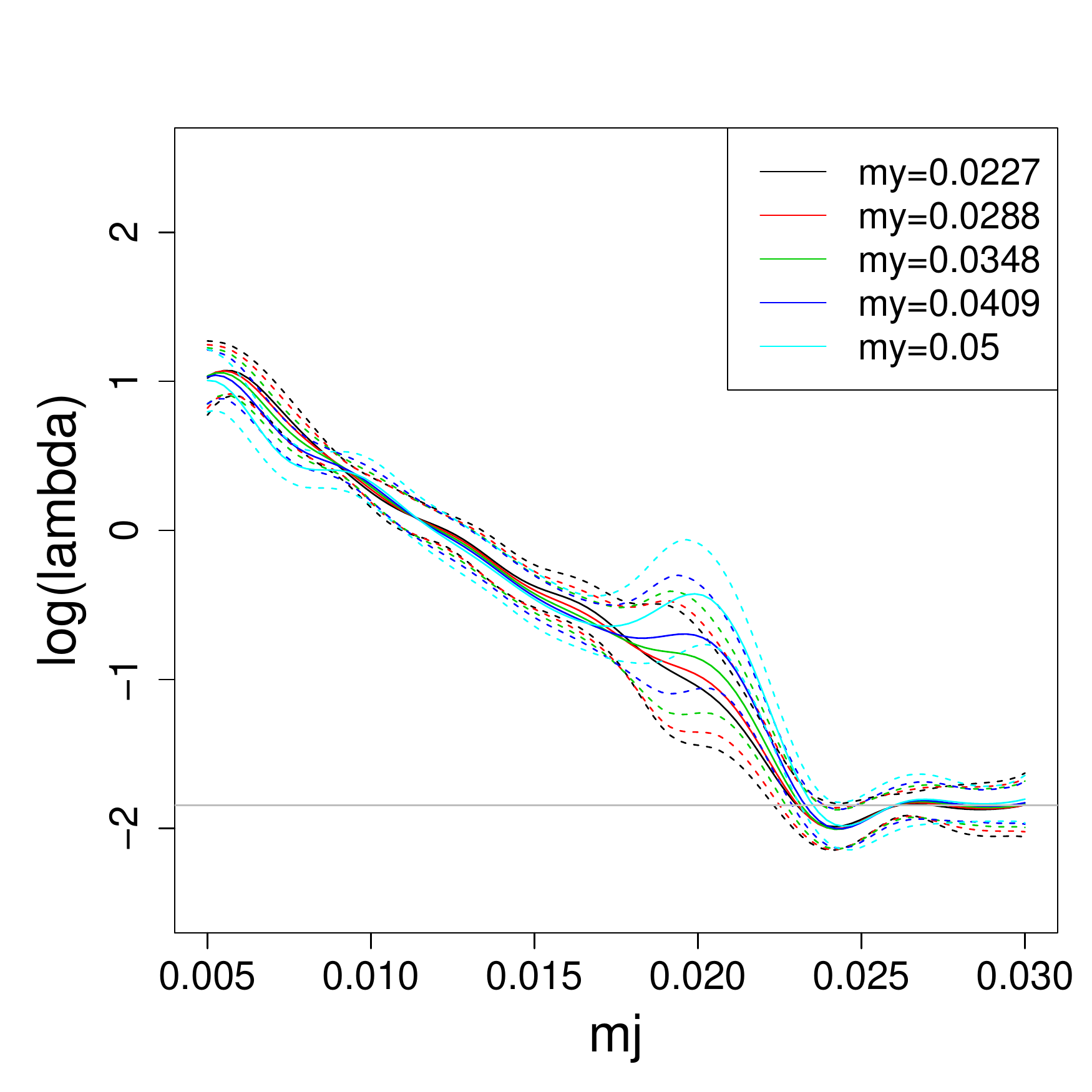}
		
	\end{tabular}
	\vspace{-0.25cm}
	\caption{Slices for the ``full'' experiment, updating Figure \ref{fig:screen}. The horizontal gray line in the right column indicates $y = \log
		\frac{1}{2}\min_{\{i:\lambda_i > 0\}} \lambda_i$, the value assigned to
		extinction outputs. }
	\label{fig:screen_new}
\end{figure}

To illustrate, see Figure \ref{fig:screen_new} which augments mean and
standard deviation slice views first provided in Figure \ref{fig:screen}.
Here, to reduce clutter, numbers overlaid indicate the degrees of replication
on {\em only} the batch/IMSPE selections. As before, these are projections
over the other five dimensions, so the connection between variance and design
multiplicity is weak (obfuscating how uncertainty relates to the other five
inputs).  Nevertheless, multiplicity in unique runs is generally higher (more
4s--6s) in the yellow regions.  The first row of Figure \ref{fig:screen_new}
coincides with Figure \ref{fig:screen}, showing input pair $m_j \times
P_{j,3}$. Observe that, after conditioning on more data despite the larger
space, predictive bands over $m_j$ are narrower, especially at the boundaries.
The sudden widening of the dark blue predictive intervals correspond to the
yellow spot in the middle panel. The second row shows a newly selected pair
$m_y \times m_j$, replacing the flat view from Figure \ref{fig:screen} which
is still uninteresting in the ``full'' setting.  A nonlinear variance is
evident, being highest near $ m_j = 0.020$.

\subsection{Downstream analysis}

Slices are certainly not the best way to visualize a high dimensional response
surface.  Moreover, there are many possible ways to utilize the information in
a fitted surrogate.  Our intent here is not to explore that vast space in any
systematic way, but rather to illustrate potential.  Here we showcase input
sensitivity analysis as one possible task downstream of fitting and design.
That is, we seek to determine which input variables have the greatest
influence on outputs, i.e., the growth rate of the fish in this example, and
which variables (if any) interact to affect changes in the response. We
perform this analysis based exclusively on the $N=2016$ runs obtained from the
batch sequential design experiment.  We could have combined with the pilot
runs, which may have reduced variability in some parts of the input space, but
could  potentially introduce complications interpretively.  

Sensitivity analysis for GP surrogates
\citep{oakley2004probabilistic,marrel2009calculations} attempts to measure the
effect of a subset of inputs on outputs by controlling and averaging over the
compliment of inputs \citep{saltelli2000sensitivity}.  \blu{In this way, one
can furnish a meaningful low-dimensional summary of inherently
high-dimensional relationships.} \citet{gramacy2020surrogates}, Chapter 8.2,
provides a thorough summary alongside a portable implementation. We briefly
summarize salient details here for completeness.

Let $U(\x) = \prod_{k=1}^m u_k(x_k)$ denote a distribution on inputs,
indicating relative importance in the range of settings or nearby nominal
values.  We take $U$ as uniform over the study region in Table
\ref{tab:smeltfullstudy}.  So-called main effects, sometimes referred to as a
zeroth-order index, are calculated by varying one input variable while
integrating out others under $U$:
\begin{equation}
\mathrm{ME}(x_j) \equiv \mathbb{E}_{U_{-j}} \{y \mid x_j\} = 
\int\!\!\int_{\mathcal{X}_{-j}} \! y P(y \mid \x) u_{-j}(x_1, x_{j-1}, x_{j+1}, x_m) \, d \mathbf{x}_{-j} dy.
\end{equation}
Above, $P(y \mid \x) = P(Y(\x) = y)$ is the predictive distribution from a
surrogate, \blu{$\hat{f}$} say via HetGP. One may approximate this
double-integral via MC with LHSs over $U$. 
We used LHSs of size 10000 paired with a common grid over each variable $j$
involved in $\mathrm{ME}(x_j)$.  

\begin{figure}[!ht]
	\centering
	\includegraphics[scale=0.5,trim=5 35 10 20, clip=true]{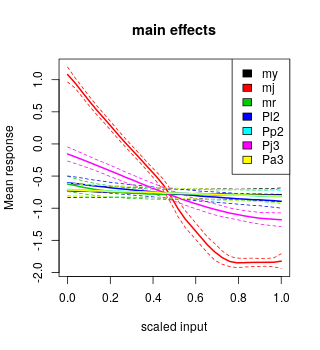}
	\includegraphics[scale=0.5,trim=20 35 10 20, clip=true]{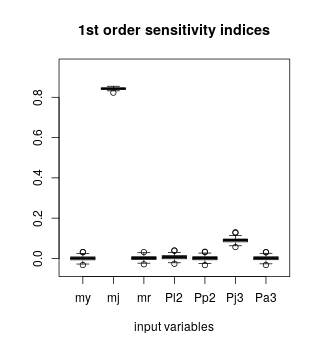}
	\includegraphics[scale=0.5,trim=10 35 10 20, clip=true]{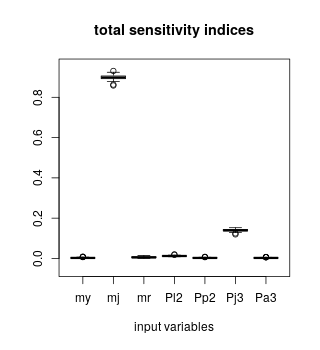}
	\includegraphics[scale=0.5,trim=0 15 10 50, clip=true]{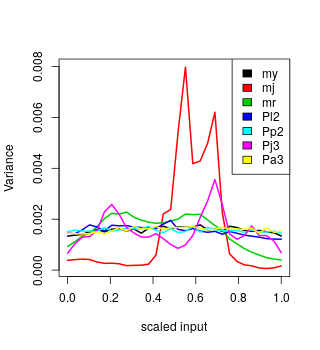}
	\includegraphics[scale=0.5,trim=20 15 10 50, clip=true]{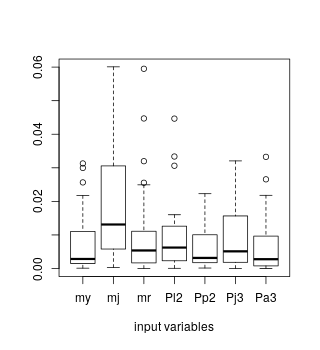}
	\includegraphics[scale=0.5,trim=10 15 10 50, clip=true]{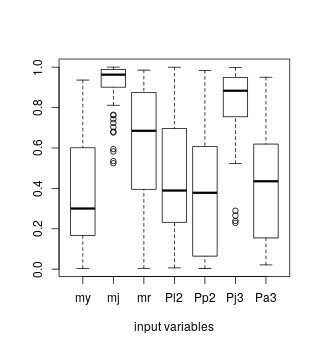}
	\caption{Sensitivity analysis for mean ({\em top}) and variance process ({\em bottom}): main effects ({\em left});  first order ({\em middle}) and total sensitivity ({\em right}) from 100 bootstrap re-samples.  }
	\label{fig:sens}
\end{figure}

The top-left of Figure \ref{fig:sens} reveals that all inputs show a negative
relationship with the response $\lambda$, with greater values leading to
declining populations.   Apparently, $m_j$ and $P_{j,3}$ induce higher mean
variation in the response than the others. These results indicate that
juvenile mortality ($m_j$) and the feeding parameter for juvenile food type 3
($P_{j,3}$) have the greatest impact on the mean value of the response (Figure
\ref{fig:sens}, top row). Further, $m_j$ exhibits a thresholding effect --
above a value of about 0.025 the simulated population almost always goes
extinct within the time frame of the simulation (Figure \ref{fig:screen_new},
right panels). However, when $m_j$ takes on values between 0.018 and 0.022, or
so, uncertainty in the estimate of mean behavior increases substantially. This
due to the individuals comprising the population becoming low in abundance and
smaller in size to the point that egg production cannot offset lifetime
mortality.

As we mentioned in Section \ref{sec:hetGP_background}, HetGP puts a second GP
prior on the latent nuggets $\Deltan$.  Once $\Deltan$ and all hyperparameters
are estimated, the predictive mean of the noise process, i.e., the smoothed
nuggets $\boldsymbol{\Lambda}$, can be calculated over any testing data set in
the domain of interest. This provides a way to assess the influence of each
input variable on the heteroskedastic variance. Applying the same procedures
as above, main effects for the noise process are produced in the bottom-left
of Figure \ref{fig:sens}. Observe that when $m_j$ is between 0.4 and 0.6
variance effects are highest, particularly for $m_j$ and $P_{j,3}$. As far as
we know, such main effects (and higher-order sensitivities) on variances are
novel in the literature.

To further quantify the variation that each input factor contributes, 
we calculated
first-order ($S$) and total ($T$) indices. These assume a functional ANOVA decomposition,
\begin{align*}
& f(x_1, \dots, x_m) = f_0
+ \sum_{j=1}^m f_j(x_j) + \!\! \sum_{1 \leq i < j \leq m} f_{ij}(x_i,x_j) +
\cdots + f_{1,\dots,m}(x_1, \dots ,x_m), \\
\mbox{so that} \quad & \mathbb{V}\mathrm{ar}_U(y \mid x_1, \dots, x_m) = \sum_{j=1}^m V_j
+ \!\! \sum_{1 \leq i < j \leq m} V_{ij} + \cdots + V_{1,\dots,m},
\end{align*}
where $V_j = \mathbb{V}\mathrm{ar}_{U_j}(\mathbb{E}_{U_{-j}}\{y \mid x_j\})$,
$V_{ij}=\mathbb{V}\mathrm{ar}_{U_{ij}}(\mathbb{E}_{U_{-ij}}\{y \mid x_i,
x_j\}) - V_i - V_j$. \blu{In a direct application, $f$ above is the simulator.
Since that is expensive in our delta smelt case, we use the surrogate
$\hat{f}$ instead. The second equation decomposes, and quantifies, variability
in  $\mathbb{E}_{U_{J}}\{y \mid \x_J\}$ with respect to changes in  $\x_{J}$
according to  $U_{J}(\x_{J})$.  It holds for our HetGP $\hat{f}$ since all
input factors can be varied independently in the input hypercube.} First-order
sensitivity $S_j$ for $x_j$ measures the proportion of variation that $x_j$
contributes to the total:
\[
S_j = \frac{\mathbb{V}\mathrm{ar}_{U_j}(\mathbb{E}_{U_{-j}}\{y \mid x_j\})}{\mathbb{V}\mathrm{ar}_U(y)}, \quad j=1,\dots,m.
\]
Total sensitivity $T_j$ is the mirror image:
\[
T_j = \frac{\mathbb{E}\{\mathbb{V}\mathrm{ar}(y \mid \mathbf{x}_{-j})\}}{\mathbb{V}\mathrm{ar}(y)} = 1-  \frac{\mathbb{V}\mathrm{ar}(\mathbb{E} \{ y \mid \mathbf{x}_{-j} \})}{\mathbb{V}\mathrm{ar}(y)}.
\]
It considers the proportion of variability that is not explained without
$x_j$. The difference between first-order and total sensitivities, i.e., $T_j
- S_j$, may be taken as a measure of variability in $y$ due to the interaction
between input $j$ and the other inputs. 

\blu{With HetGP surrogate $\hat{f}$ for $y\mid \x$}, calculation of
$S$ and $T$ indices may also be undertaken by MC via LHS. The details are
omitted here for brevity. We repeated  MC calculations of both on 100
bootstrap samples of the original data set.  A summary via boxplots is
provided in the right panels of Figure \ref{fig:sens}. These views match the
main effects: $m_j$ and $P_{j,3}$ stand out in both plots. First-order
sensitivity ($S$) for the variance (bottom-middle), fails to flag an obvious
difference between variables. Via total sensitivity ($T$, bottom-right),
indices for $m_j$ and $P_{j,3}$ are again apparently higher than other
variables, suggesting that a substantial aspect of the effect of these
variables on variability is through interactions.

\begin{table}[h!]
	\centering
	\begin{tabular}{l| c c c c c c c}
		Proportion& $m_y$ & $m_j$ &$m_r$ &$P_{l,2}$ &$P_{p,2}$ &$P_{j,3}$ &$P_{a,3}$\\
		\hline 
		Main process & 0.54 &1 & 0.66 &0.68 &0.53 &1 &0.55\\
		Noise process & 0.74 &0.99 & 0.92 &0.80 &0.75 &0.97 &0.80
	\end{tabular}
	\caption{Proportion of positive $I=T-S$ indices.}
	\label{tab:T-S}
\end{table}

Using those $S$ and $T$ values, we computed $I=T-S$.  \blu{As
\citet{saltelli2000sensitivity} describes, this quantity $I$ is positive if
variability in the response can be attributed to interactions between inputs.}
The proportion of our bootstrapped $I$ measurements which are positive is
provided in Table \ref{tab:T-S}.  Again, $m_j$ and $P_{j,3}$ flag has highly
probable for impacting the response through an interaction with other
variables. Input $P_{l,2}$ may also have substantial impact on $\lambda$
through interactions. Input $m_r$ has the third highest measurement. In fact,
all of the variables suggest statistically noteworthy affect through
interaction by comparison to the so-called {\em median probability model}
\citep{barbieri2004optimal} implied by a $p=0.5$ threshold.  Since not many
variables contribute via zeroth (main effect) and first-order summaries, it is
perhaps not surprising that action is exposed through interaction.

The results of this analysis reveal that although juvenile mortality is
important for the overall mean, it by itself does not determine whether a
population will increase or decrease (i.e., $\lambda>0$ or $< 0$). For
example, although it will typically be important for the average value of
$m_j$ to be \blu{below} 0.015 or so, the average value of $\lambda$ will
depend, sometimes sensitively, on the values of other parameters (Figure
\ref{fig:screen_new}, top right panel), or through their interactions (Table
\ref{tab:T-S}).  \blu{The complex nature of these sensitivities implies that
environmental variation and management actions that affect these key
parameters will likely generate non-linear responses in population growth rate
that depend on the interactive effects among parameters.  In addition,
several parameters (e.g., $m_r$) have important effects on the variance of
population growth rate, further complicating any assessment of the value of an
individual parameter.  Predicted population responses to actions directed at
affecting change in specific biological processes (e.g., mortality rate of
juveniles, $m_j$) should be viewed in the context of the state of the system
and population. Simple changes to key parameters will likely not lead to
simple responses (i.e., change in magnitude and variance of growth rate) but
rather to ones that depend on the values of other parameters.}

More importantly than the specific predictions, however, the analysis allows
us to characterize the emergent behavior of this complex simulation model. The
goal of ABMs is to incorporate possible mechanisms that are believed to impact
dynamics of complex systems. However model builders cannot typically predict,
a priori, the full dynamics of the model across all possible parameter
settings. An appropriately designed surrogate allows us to quickly probe the
model more deeply without needing to re-run the full simulation at every
parameter combination of interest, potentially at enormous computational
expense.  \blu{For example, by independently varying input factors, we are
able to compare their relative contribution and investigate
higher-order interactions without new runs of the simulator.} Further, once
dominant variables are identified, fixing them can unravel the sensitivities
of the remaining variables. As one further example, we fixed $m_j$ and
$P_{j,3}$ and made an analogue of Figure \ref{fig:sens} over the other five
factors.  See Appendix \ref{app:sens} for details.

\section{Discussion}
\label{sec:futurework}

Motivated by a computationally intensive stochastic agent-based model
simulating the ecosystem and life cycles of delta smelt, an endangered fish,
we developed a batch sequential design scheme for loading supercomputing nodes
with runs in batches.  We used a heteroskedastic Gaussian process (HetGP)
surrogate to acknowledge nonlinear dynamics in mean and variance,
revealed in a limited pilot study, and extended a variance-based (IMSPE)
scheme for sequential design under such models to allow the selection of
multiple new runs at once.  To facilitate numerical optimization of batch
IMSPE we furnished closed form derivatives and developed a backtracking scheme
to determine if any near replicates provided by the solver were better as actual
replicates.  Only actual replicates efficiently separate signal from noise and
pay computational dividends at the same time.

Our methods were illustrated and contrasted against previous (pure
sequential/one-at-a-time) active learning strategies on several synthetic and
real-simulation benchmarks.  These allowed us to conclude that our scheme was
no worse than previous approaches, while designing batches of runs that could
fill out a supercomputing node.  \blu{Since those one-at-a-time schemes
already outperformed single-batch designs, we conclude by proxy that ours do
as well, and verified as much in additional experimentation (not provided).}
We then turned to our motivating delta smelt scenario to undertake a
simulation campaign  with thousands of runs in an expanded domain.
\blu{Those simulations required 12000 core hours, which would have spanned
more than 500 days if run back-to-back (and not counting any queue delays).}
\blu{Instead the batch campaign, took us about 55 days to run (including
substantial queuing).}

This order of magnitude reduction in ``scientist time'', without noticeable
drawbacks in modeling efficiency,  could have a substantial impact on the
modus operandi of conducting stochastic simulation experiments in practice.
Widespread university and research lab access to supercomputing facilities is
democratizing the application of mathematical modeling of complex physical and
biological phenomena.  However, strategies for planning those experiments in
this unique architectural environment are sorely needed.  We think the
advances reported on here take an important first step. Simulations in hand,
there are many interesting analyses which can be performed downstream.  We
provided some visuals based on slices and performed an input sensitivity
analysis in order to determine which factors have the largest effect on smelt
mortality in this particular system.  Our choice of IMSPE suits this analysis
well because it reduces variance globally and our Saltelli-style indices
emphasize decomposition of variance.  Extending \citeauthor{HetGP2}'s IMSPE
calculation to other downstream tasks has become a cottage industry of late.
Examples include sequential learning of active subspaces
\citep{wycoff2019sequential}, level-set finding and Bayesian optimization
\citep{lyu2018evaluating}.  \citet{cole2020locally} adapt a similar
calculation for large-scale local GP approximation via inducing
points.  We see no barriers to extending these schemes similarly, to batch
analogs of one-at-a-time acquisitions. \citet{kenn:ohag:2001}-style
calibration of stochastic simulators remains on the frontier of design for
surrogate modeling. \citet{baker2020stochastic} identify this as an important
area for further research.

\blu{One might wonder if a Binois-like one-at-a-time scheme could be adapted for
batch acquisition by inserting ``synthetic data'', obtained from the
predictive distribution,  for earlier (waiting to run) selections while
entertaining the selection of latter batch elements. Although there are many
ways to operationalize such an idea, we think this is flawed for several
reasons.  One is that the high and changing noise scenario would demand
entertaining large numbers of replicates for those synthetic data, eliminating
any computational advantage.  We found this too prohibitive to entertain as a
comparator. Another is that such a scheme would be unnecessarily making a
greedy approximation to a joint optimization. Although joint derivatives are
much more work to derive, they are simple to code now that we have provided
them. No greedy approximation is necessary.}
	
There is certainly potential for improving our scheme. \blu{We took a
geometric mean to get scalar output for smelt simulations. Analyzing
higher-dimensional outputs could reveal additional nuance, however HetGP
surrogate modeling for such settings remains on the frontier.}  The
performance of our scheme relies heavily on local numerical optimization via
libraries. Finding global optima for non-convex criteria in high-dimensional
spaces is always a challenge.  Although we get good results with L-BFGS-B, we
also tried particle swarm optimization \citep[PSO;][]{PSO} in several
capacities: replacing BFGS wholesale and for finding good BFGS staring points.
Improvements were consistent but minor in the grand scheme of multiple batches
of sequential design. \blu{We believe that other gradient-free/coordinate
exchange methods would perform similarly.} Hybrid genetic and gradient-based
optimization \citep{Mebane2011GeneticOU} could be promising, as could a
weighted least-squares approach to identifying candidate numbers of replicates
\citep{deng:2018}.
\blu{Trade-offs between queue time and batch size might be worth
exploring, however that could be a fractal undertaking:  a meta computer
experiment would be needed to map out the response surface of run/wait times
based on run configuration and other (computing) environment variables.}  We
kept it simple with $M=24$ to match the number of cores on the compute nodes,
\blu{as advised by the VT ARC office.} Another option is unknown batch sizes
or on-demand acquisition: whenever a batch of cores is available the
model/design scheme must be ready to furnish runs.  This could be accomplished
by maintaining a larger $M$-sized queue of prioritized inputs, say following
\citet{gra:lee:2009}, which would need to be updated for the HetGP framework.

Focusing on the delta smelt ABM in particular, the current version of the
simulator makes several fundamental assumptions that influence the population
dynamics and can affect measured input sensitivities. Future analysis could
accommodate additional parameters, such as those related to juvenile and adult
movement behaviors that affect their growth and mortality. One could also
address structural uncertainty in the delta smelt ABM.  For example, movement
rates are held constant for all individuals within each life stage and
maturity is a fixed threshold function of length.  \cite{kenny:2013_2}
examined alternative setups to assess how different assumptions would affect
model results. They formulated mortality to continuously decrease with length,
to be density-dependent (rather than constant), and substituted a smoothed
function of maturity-by-length for the threshold, and repeated the ten year
simulations. A future simulation campaign (perhaps using the same parameters
selected for the analysis reported here or under a novel batch-sequential
design) repeated under these alternative assumptions would provide valuable
additional information on model sensitivities and uncertainties.

\subsubsection*{Acknowledgments} 

Authors BZ and RBG gratefully acknowledge funding from a DOE LAB 17-1697 via
subaward from Argonne National Laboratory for SciDAC/DOE Office of Science
ASCR and High Energy Physics. RBG recognizes partial support from National
Science Foundation (NSF) grant DMS-1821258. LRJ recognizes partial support from NSF grant DMS/DEB-1750113.  We also gratefully acknowledge
computing support from Virginia Tech's Advanced Research Computing (ARC)
facility. We thank Xinwei Deng, Dave Higdon and Leanna House (Virginia Tech) for valuable insights and suggestions.

\bibliographystyle{jasa}
\bibliography{smelt}

\appendix
	
\section{IMSPE gradient}
\label{sec:Igrad}

Omitted expressions for the IMSPE gradient in Section \ref{sec:gradient} are
provided below. 
\begin{align*}
\dfrac{ \partial \K^{-1}_{n+M}}{\partial \xnewip} & =\dfrac{ \partial} {\partial \xnewip} 
\begin{bmatrix}
\K_{n}^{-1} + g(\xnew) \Sigma(\xnew) g(\xnew)^ \top  & g(\xnew) \\
g(\xnew) ^\top & \Sigma(\xnew)^{-1}
\end{bmatrix}
= \begin{bmatrix}
H(\xnew) & Q(\xnew) \\
Q(\xnew)^\top & V(\xnew)
\end{bmatrix} \quad , \\
\mbox{where} \;\; V(\xnew) &:=
-\Sigma(\xnew)^{-1} \frac{\partial \Sigma(\xnew)}{\partial \xnewip} \Sigma(\xnew)^{-1} \nonumber \\
Q(\xnew) & := \dfrac{ \partial g(\xnew) } {\partial \xnewip}
= - \K_n^{-1} \left(c(\vecXu, \xnew) V(\xnew) + \textcolor{black}{\frac{\partial c(\vecXu, \xnew)}{\partial \xnewip}} \Sigma(\xnew)^{-1}  \right) \nonumber \\
H(\xnew) &:=\dfrac{ \partial g(\xnew) \Sigma(\xnew) g(\xnew)^ \top } {\partial \xnewip}\\
&=  g(\xnew) \dfrac{ \partial \Sigma(\xnew) } {\partial \xnewip} g( \xnew)^\top
\!+\! Q(\xnew) \Sigma(\xnew) g(\xnew)^\top\!+\!\{Q(\xnew) \Sigma(\xnew) g(\xnew)^\top\}^\top. \nonumber 
\end{align*}

Recall that $\Sigma(\xnew) = r(\xnew) + c(\xnew, \xnew) - c(\vecXu,
\xnew)^\top \K_{n}^{-1}c(\vecXu, \xnew)$. Again recursing with the chain
rule, first 
diagonal matrix $r(\xnew)$ via
Eq.~\eqref{eq:rhat}, gives
\begin{equation}
\frac{\partial r(\xnew)}{\partial \xnewip} 
=  \frac{\partial \blu{c}_{(\boldsymbol{\delta})}(\xnew, \vecXu) }{\partial \xnewip} (\Cdelta + g_{(\boldsymbol{\delta})}\A^{-1})^{-1} \Deltan.
\label{equ:dr}	
\end{equation}
It is worth observing here how relative noise levels, smoothed through
$\Deltan$ and distance to $\vecXu$, impact the potential value of new design
elements $\xnew$.  High variance $\bar{\mathbf{x}}_i$ have low
impact unless $a_i$ is also large, in which case there is an attractive force
encouraging replication (elements of $\xnew$ nearby $\vecXu$). 
The last component of $\frac{\partial
	\Sigma(\xnew)}{\partial \xnewip}$ relies on $\frac{\partial c(\vecXu,
	\xnew)}{\partial \xnewip}$, a quadratic:
\begin{align}
\frac{\partial}{\partial \xnewip} & c(\vecXu, \xnew)^\top \K_{n}^{-1}c(\vecXu, \xnew) \label{equ:dc} \\
&= c(\vecXu, \xnew)^\top \K_{n}^{-1}\frac{\partial c(\vecXu, \xnew)}{\partial \xnewip} 
+ \left\{c(\vecXu, \xnew)^\top \K_{n}^{-1}\frac{\partial c(\vecXu, \xnew)}{\partial \xnewip}\right\}^\top.
\nonumber
\end{align}
The structure of this component's derivative reveals how new design elements
$\xnew$ repel one another and push away from existing points $\vecXu$. In
other words, the forces described in Eqs.~(\ref{equ:dr}--\ref{equ:dc})
trade-off in a sense, encouraging both spread to space-fill and compression
toward replication depending on the noise level $r(\cdot)$.

The other terms that are included in $\frac{\partial \Sigma(\xnew)}{\partial \xnewip}$ are as follows:
\begin{align*}
\frac{\partial c(\vecXu, \xnew)}{\partial \xnewip} & = \frac{\partial}{\partial \xnewip} \begin{bmatrix}
c(\vecxnew_1, \xu_1) & \!\cdots\! & c(\vecxnew_M, \xu_1)\\
\vdots & & \vdots \\
c(\vecxnew_1, \xu_n) & \!\cdots\! & c(\vecxnew_M, \xu_n)\\ 
\end{bmatrix} = \begin{bmatrix}
& \frac{\partial c(\vecxnew_i, \xu_1)}{\partial \xnewip} &\\
\mathbf{0}_{n \times (i-1)} & \vdots & \mathbf{0}_{n \times (M-i)}\\
& \frac{\partial c(\vecxnew_i, \xu_n)}{\partial \xnewip} &\\
\end{bmatrix} \nonumber \\
\frac{\partial c(\xnew, \xnew)}{\partial \xnewip} & = \frac{\partial}{\partial \xnewip} 
\begin{bmatrix}
c(\vecxnew_1, \vecxnew_1) & \!\!\!\cdots\!\!\! & c(\vecxnew_M, \vecxnew_1)\\
\vdots & & \vdots\\
c(\vecxnew_1, \vecxnew_M) & \!\!\!\cdots\!\!\! & c(\vecxnew_M,\vecxnew_M)\\ 
\end{bmatrix} = \begin{bmatrix}
&  & \frac{\partial c(\vecxnew_i, \vecxnew_1)}{\partial \xnewip} & & \\
&  & \vdots & &  \\
\frac{\partial c(\vecxnew_i, \vecxnew_1)}{\partial \xnewip} & \!\!\cdots\!\! & \frac{\partial c(\vecxnew_i, \vecxnew_i)}{\partial \xnewip} & \!\!\cdots\!\! & \frac{\partial c(\vecxnew_i, \vecxnew_M)}{\partial \xnewip}\\
&  & \vdots & & \\
&  & \frac{\partial c(\vecxnew_i, \vecxnew_n)}{\partial \xnewip}& &\\ 
\end{bmatrix}.
\end{align*}

To ensure positive variances, i.e., rather than being faithful to Eq.~(\ref{equ:rhatpred}), we 
instead model
\begin{align*}
\log r(\xnew) &=  
\blu{c}_{(\boldsymbol{\delta})}(\xnew, \vecXu)(\Cdelta + g_{(\boldsymbol{\delta})} \An^{-1})^{-1} \log \Deltan.
\end{align*}
\blu{Here, $\log \Deltan$ is optimized jointly.} Thus $\frac{\partial
r(\xnew)}{\partial \xnewip}$ can be derived as:
\begin{align*}
\frac{\partial r(\xnew)}{\partial \xnewip} 
& =  \frac{\partial K_{(\boldsymbol{\delta})}(\xnew, \vecXu) }{\partial \xnewip} (\Cdelta + g_{(\boldsymbol{\delta})}\A^{-1})^{-1} \log \Deltan \\
& \quad \quad \times \exp(K_{(\boldsymbol{\delta})}(\xnew, \vecXu) (\Cdelta + g_{(\boldsymbol{\delta})}\A^{-1})^{-1} \log \Deltan)
\end{align*}
Then we focus on the expressions related to $\dfrac{ \partial \W_{n+M}}{\partial \xnewip}$.	
\begin{align*}
\dfrac{\partial\W_{n+M}}{\partial \xnewip} &
= \dfrac{\partial}{\partial \xnewip} \begin{bmatrix}
\Wn & w(\vecX_n, \xnew) \\
w(\vecX_n, \xnew)^{\top} & w(\xnew, \xnew)
\end{bmatrix}
= \begin{bmatrix}
\mathbf{0} & S(\xnew)\\
S(\xnew)^{\top} & T(\xnew)
\end{bmatrix}, \quad 
\mbox{where} \\
S(\xnew) & = \frac{\partial}{\partial \xnewip} \begin{bmatrix}
w(\vecxnew_1, \xu_1) & \dots & w(\vecxnew_M, \xu_1)\\
\vdots & & \vdots\\
w(\vecxnew_1, \xu_n) & \dots & w(\vecxnew_M, \xu_n)\\ 
\end{bmatrix} = \begin{bmatrix}
& \frac{\partial w(\vecxnew_i, \xu_1)}{\partial \xnewip} &\\
\mathbf{0}_{n \times (i-1)} & \vdots & \mathbf{0}_{n \times (M-i)}\\
& \frac{\partial w(\vecxnew_i, \xu_n)}{\partial \xnewip} &\\
\end{bmatrix} \nonumber \\
T(\xnew) & = \frac{\partial}{\partial \xnewip} \begin{bmatrix}
w(\vecxnew_1, \vecxnew_1) & \dots & w(\vecxnew_M, \vecxnew_1)\\
\vdots & & \vdots\\
w(\vecxnew_1, \vecxnew_M) & \dots & w(\vecxnew_M,\vecxnew_M)\\ 
\end{bmatrix} = \begin{bmatrix}
&  & \frac{\partial w(\vecxnew_i, \vecxnew_1)}{\partial \xnewip} & & \\
&  & \vdots & &  \\
\frac{\partial w(\vecxnew_i, \vecxnew_1)}{\partial \xnewip} & .. & \frac{\partial w(\vecxnew_i, \vecxnew_i)}{\partial \xnewip} & .. & \frac{\partial w(\vecxnew_i, \vecxnew_M)}{\partial \xnewip}\\
&  & \vdots & & \\
&  & \frac{\partial w(\vecxnew_i, \vecxnew_n)}{\partial \xnewip}& &\\ 
\end{bmatrix}. \nonumber
\end{align*}
For Gaussian kernel, $w(\cdot, \cdot)$ is calculated with erf the error function $\erf (z) = \frac{2}{\sqrt{\pi}} \int \limits_{0}^z e^{-t^2}\, dt$ as
\begin{align*}
w(x_i, x_j) =& \frac{\sqrt{2\pi\theta}}{4} \exp\left(- \frac{(x_i-x_j)^2}{2\theta}\right) \left(\erf \left(\frac{2 - (x_i + x_j)}{\sqrt{2 \theta}}\right) + \erf \left(\frac{x_i + x_j}{\sqrt{2 \theta}}\right)\right), 
\end{align*}
for $1 \leq i,j \leq n$ and with derivative
\begin{multline*}
	\frac{\partial w(x, x_i)}{\partial x} =
	\sqrt{\frac{\pi}{8 \theta}} \exp \left( -\frac{(x-x_i)^2}{2 \theta}\right) \left[ (x-x_i)
	\left( \erf \left(\frac{x + x_i - 2}{\sqrt{2 \theta}}\right) - \erf \left(\frac{x + x_i}{\sqrt{2 \theta}}\right)\right) \right. \\
	\left. + \sqrt{\frac{2\theta}{\pi}} \left( \exp \left( - \frac{(x+x_i)^2}{2\theta}\right) - 
	\exp \left( \frac{-(x + x_i -2)^2}{2\theta}\right)\right) \right].
	\end{multline*} 

\section{More examples}
\label{app:examples}

Here we describe the two example omitted from Section \ref{sec:examples}.

\subsection{1d toy example}

This 1d synthetic example was introduced by \citet{HetGP2} to show how IMSPE
acquisitions distribute over the input space in heteroskedastic settings. Here
we borrow that setup to illustrate our batch scheme. The underlying true mean
function is $f(x) = (6x - 2)^2 \sin(12x - 4)$, and the true noise function is
$r(x) = (1.1 + \sin(2 \pi x))^2$. Observations are generated as $y \sim f(x) +
\epsilon$, where $\epsilon \sim N(0, \sigma^2 = r(x))$. The experiment starts
with a maximin--LHS of $n_0=12$ locations under a random number of replicates
uniform in \{1, 2, 3\}, so that the starting size is about $N_0=24$. A total
of twenty $M=24$-sized batches are used to augment the design for a total
budget of $N=504$ runs.

\begin{figure}[ht!]
\centering
\includegraphics[scale=0.34,trim=0 60 10 20,clip=TRUE]{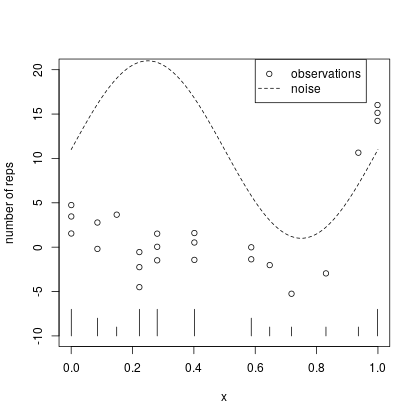}
\includegraphics[scale=0.34,trim=50 60 10 20,clip=TRUE]{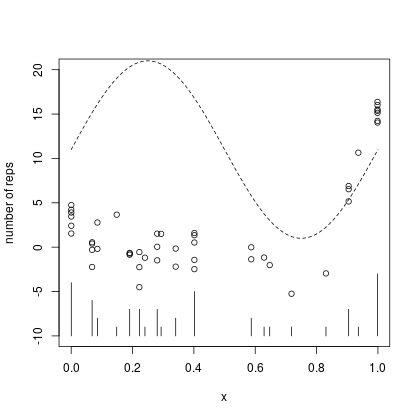}
\includegraphics[scale=0.34,trim=50 60 10 20,clip=TRUE]{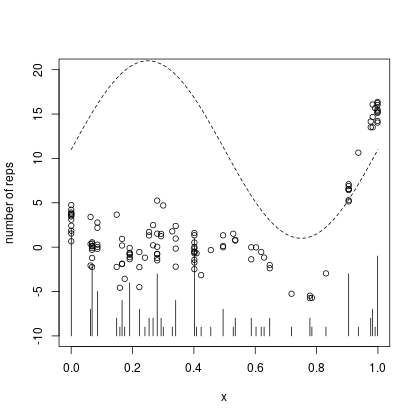}
\includegraphics[scale=0.34,trim=0 20 10 40,clip=TRUE]{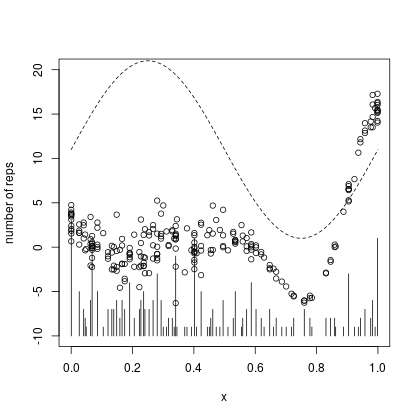}
\includegraphics[scale=0.34,trim=50 20 10 40,clip=TRUE]{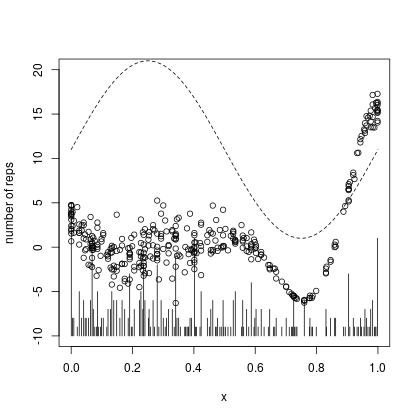}
\includegraphics[scale=0.34,trim=50 20 10 40,clip=TRUE]{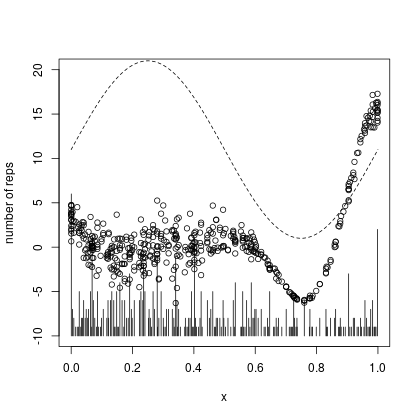}

\caption{The top-left panel shows the initial design observations. Remaining
panels display the sequential design process after adding 1, 5, 10, 15 and 20
batches.}
\label{fig:1d}
\end{figure}

Panels in Figure \ref{fig:1d} serve to illustrate this process in six
epochs. Open circles indicate observations, with more being added in batches
over the epochs. The dashed sine curve indicates the relative noise level
$r(x)$ over the input space; vertical segments at the bottom highlight the
degree of replication at each unique input. Observe how more runs are added to
high noise regions, and the degree of replication is higher there too.  This
is strikingly similar to the behavior reported by \citeauthor{HetGP2}.

\subsection{Ocean oxygen}

\begin{figure}[ht!]
\centering
\includegraphics[width=0.96\textwidth]{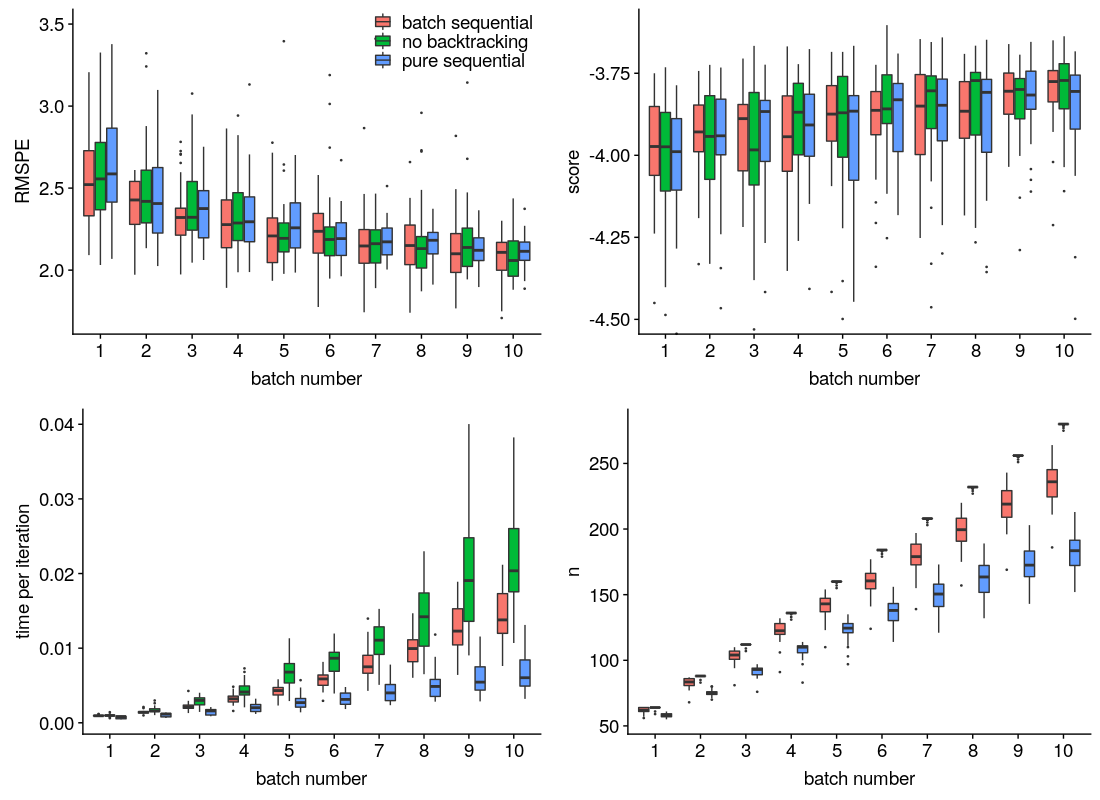}
\caption{Ocean simulator results in 30 MC repetitions: RMSPE, score, 
time per batch and the aggregate number of unique design locations $n$.}
\label{fig:ocean}	
\end{figure}

The ocean-oxygen simulator models oxygen concentration in a thin water layer
deep in the ocean, see \citet{ocean}. For details on how we generate
simulations here, see \citet{Herbei2014estimating} and \citep[][Section
10.3.4]{gramacy2020surrogates}.\footnote{Implementation is provided
\url{https://github.com/herbei/FK_Simulator}.}  The simulator is stochastic
and is highly heteroskedastic. Visuals are provided by our references
above.  There are four real-valued inputs: two spatial coordinates (longitude
and latitude) and two diffusion coefficients.  We consider a MC experiment
initialized a $n_0 = 40$-sized maximin--LHS, with five replicates upon each
($N_0 = 200$). We consider adding ten $M=24$-sized batches so that $N=440$ runs
are collected by the end.   We can't easily visualize the results in a 4d
space, but the analog of our 2d toy results (Figure \ref{fig:2dresult}) is
provided in Figure \ref{fig:ocean}.

In terms of out-of-sample RMSPE and score, all methods exhibit similar
performance. The purely sequential design method consistently yields more
replicates. Thus, it also takes the lowest time per iteration for updating via
{\tt hetGP}. Our backtracking scheme yields a moderate proportion of
replicates with the same performance as measured by RMSPE and score, compared
to the version without backtracking. Notice that these metreics do not
necessarily improve monotonically over batches.  This could be attributed to
unknown ``true'' mean and noise functions in this real-world simulator
setting. Calculation of RMSPE and score are out-of-sample, on novel random
testing sets, interjecting an extra degree of stochasticity in these
assessments.

\section{Sanity check on batch acquisition} 
\label{app:dist}

\begin{figure}[!ht] 
\centering
\includegraphics[scale=0.52,trim=10 15 10 28]{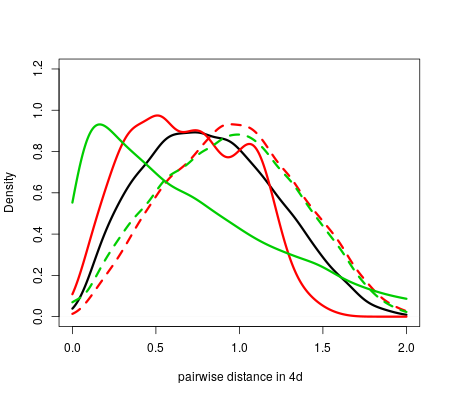}
\includegraphics[scale=0.52,trim=10 15 10 28]{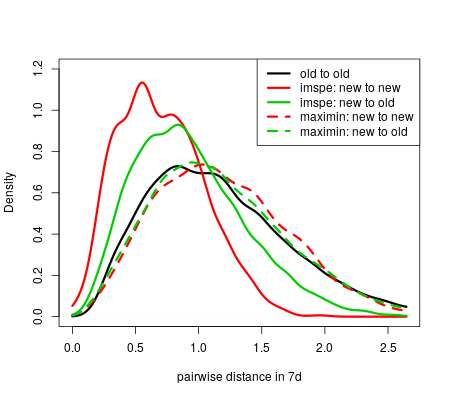}
\caption{Empirical density of pairwise distances from IMSPE batch and maximin
sequential design for the pilot (left) and full (right) studies.}
\label{fig:pairwise_density}
\end{figure}

We returned to the 6d (4d manifold) pilot study of Section \ref{sec:pilot},
involving $N=480$ runs, and inspected the properties of two new $M=24$-sized
batches. To understand how these 48 inputs, selected via IMSPE and
backtracking based on HetGP, compare to the original $n=96$-sized
space-filling design,  we plotted empirical densities of pairwise Euclidean
distances within and between the two sets.  See the solid-color-lined
densities in the left panel of Figure \ref{fig:pairwise_density}. Dashed
analogues offer a benchmark via sequential maximin design in two similarly
sized batches.  These represent an alternative, space-filling default,
ignoring HetGP model fit/IMSPE acquisition criteria.\footnote{Sequential
maximin, being model-free, doesn't require new evaluations of the simulator.}

Consider first comparing the solid and dashed green lines, capturing the
spread of distances between new and old runs.  Observe that the solid-green
density is shifted to the left relative to the dashed ones.  This view reveals
that IMSPE-selected runs are closer to the existing ones than they
would be under a space-filing design.  The solid-green density is similarly
shifted left compared to distances from the old space-filling design
(solid-black).  The situation is a little different for distances within the
new batches shown in red.  Here we have a tighter density for IMSPE compared
to space-filling, meaning we have fewer short and long distances -- more
medium ones.  We take this as evidence that the HetGP/IMSPE batch scheme
is working: spreading points out to a degree, but also focusing on some
regions of the input space more than others.  

The right panel of Figure \ref{fig:pairwise_density} shows an analog of the
comparison of pairwise distances for the larger smelt simulation campaign
described in Section \ref{sec:deltasmelt}.  In those 1056 new acquisitions,
204 involved replicates. With many more distance pairs, these kernel densities
are more stable than in the 4d case on the left. Nonetheless, we observe a
similar pattern here in 7d.  IMSPE selections tend to be closer to themselves
and to existing locations than ordinary space-filling ones would.  We take
this as an indication that the scheme was acting in a non-trivial way to
reduce predictive uncertainty captured by HetGP model fits.

\section{Sensitivity indices for $m_r$}
\label{app:sens}

As we introduced in Section \ref{sec:abm}, $m_r$ is a parameter that describes
the influence of water diversion facilities on $\lambda$. Here, we fix the two
most influential factors, $m_j$ and $P_{j,3}$, at their respective default
values of 0.015 and 0.6, to investigate sensitivity indices (based on our full
analysis from Section \ref{sec:deltasmelt}) for other input factors.
\begin{figure}[!ht]
	\centering
	\includegraphics[scale=0.41,trim=5 35 10 20, clip=true]{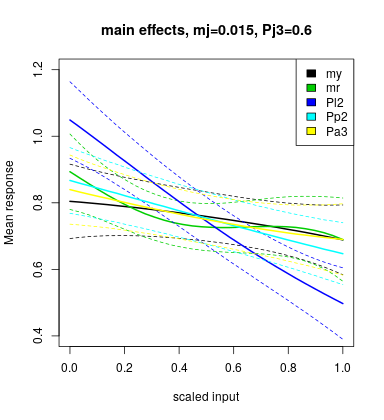}
	\includegraphics[scale=0.415,trim=20 35 10 20, clip=true]{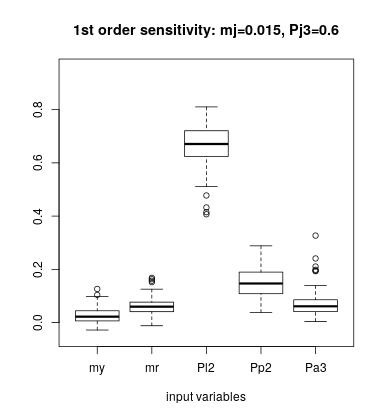}
	\includegraphics[scale=0.41,trim=10 35 10 20, clip=true]{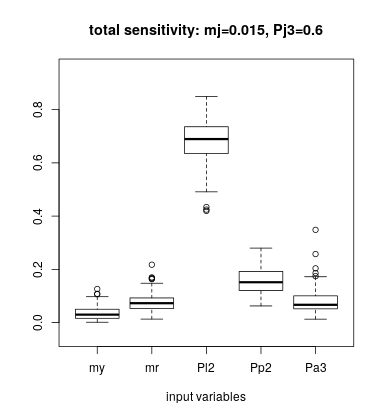}
	\includegraphics[scale=0.41,trim=0 15 10 50, clip=true]{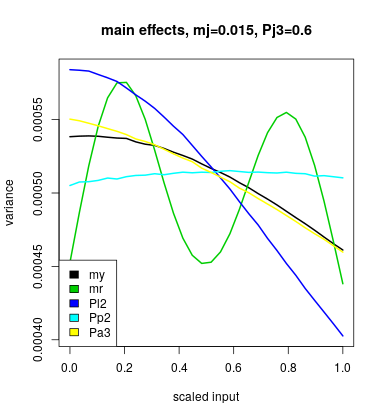}
	\includegraphics[scale=0.41,trim=20 15 10 50, clip=true]{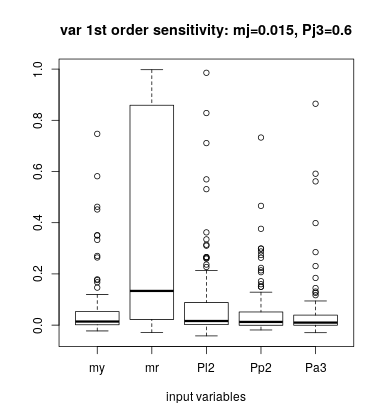}
	\includegraphics[scale=0.41,trim=10 15 10 50, clip=true]{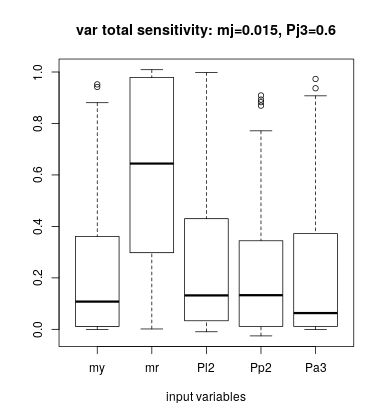}
	\caption{Sensitivity analysis for mean ({\em top}) and variance process ({\em bottom}) with  $m_j$ and $P_{j,3}$ fixed: main effects ({\em left});  first order ({\em middle}) and total sensitivity ({\em right}) from 100 bootstrap re-samples.  }
	\label{fig:sens_fix_mj_Pj3}
\end{figure}
Observe in the top left panel of Figure \ref{fig:sens_fix_mj_Pj3} that,
besides $m_j$ and $P_{j,3}$, \blu{$P_{l,2}$} also results in a relatively high
variability of the mean response.  Since the green curves are nonlinear, our
analysis shows that $m_r$ plays a complicated role in the simulated population
dynamics. Bottom panels of the figure show that $m_r$
settings drive variation in noise more than the remaining five input factors.

\end{document}